\documentclass{article}
\usepackage{arxiv}
\usepackage{url}
\usepackage{multirow}
\usepackage{textcomp}
\usepackage{algorithm}
\usepackage[noend]{algpseudocode}
\usepackage{graphicx}
\graphicspath{{figure/}}
\usepackage{epstopdf}
\usepackage{subfloat}
\usepackage{subfig}
\usepackage{amsmath}
\usepackage{multicol}
\usepackage{amssymb}
\usepackage{multicol}
\usepackage{graphicx}
\usepackage[utf8]{inputenc} % allow utf-8 input
\usepackage[T1]{fontenc}    % use 8-bit T1 fonts
\usepackage{hyperref}       % hyperlinks
\usepackage{url}            % simple URL typesetting
\usepackage{booktabs}       % professional-quality tables
\usepackage{amsfonts}       % blackboard math symbols
\usepackage{nicefrac}       % compact symbols for 1/2, etc.
\usepackage{microtype}      % microtypography
\usepackage{lipsum}

\def\BibTeX{{\rm B\kern-.05em{\sc i\kern-.025em b}\kern-.08em
    T\kern-.1667em\lower.7ex\hbox{E}\kern-.125emX}}
 \providecommand{\keywords}[1]{\textbf{\textit{Index terms---}} #1}

\title{ Digital Forensics vs. Anti-Digital Forensics: Techniques, Limitations and Recommendations}
\author{Jean-Paul A. Yaacoub, Hassan N. Noura, Ola Salman and Ali Chehab\\American University of Beirut,\\Electrical and Computer Engineering Department,\\ Beirut 1107 2020, Lebanon\\
}

\begin{document}
% \tableofcontents
\maketitle

\begin{abstract}
The number of cyber attacks has increased tremendously in the last few years. This resulted into both human and financial losses at the individual and organization levels. Recently, cyber-criminals are leveraging new skills and capabilities by employing anti-forensics activities, techniques and tools to cover their tracks and evade any possible detection. Consequently, cyber-attacks are becoming more efficient and more sophisticated. Therefore, traditional cryptographic and non-cryptographic solutions and access control systems are no longer enough to prevent such cyber attacks, especially in terms of acquiring evidence for attack investigation. Hence, the need for well-defined, sophisticated, and advanced forensics investigation tools are highly required to track down cyber criminals and to reduce the number of cyber crimes. This paper reviews the different forensics and anti-forensics methods, tools, techniques, types, and challenges, while also discussing the rise of the anti-anti-forensics as a new forensics protection mechanism against anti-forensics activities. This would help forensics investigators to better understand the different anti-forensics tools, methods and techniques that cyber criminals employ while launching their attacks. Moreover, the limitations of the current forensics techniques are discussed, especially in terms of issues and challenges. Finally, this paper presents a holistic view from a literature point of view over the forensics domain and also helps other fellow colleagues in their quest to further understand the digital forensics domain.
\end{abstract}


%\begin{multicols}{2}

\section{Introduction}
In cyber forensics, investigators aim to retrieve digital evidence from digital and cyber/physical devices including network devices, computers, smart and mobile sensors and devices, as well as drones and robots. Unfortunately, forensics investigations are not very effective due to the increasing use of anti-forensics techniques. In fact, current forensics approaches suffer from different technical flaws due to the anti-forensics tools to avoid detection. The anti-forensics techniques are used to disable and distort forensics investigation by attacking the forensics tools or by deleting, hiding or encrypting the evidence itself. More specifically, some anti-forensics tools are used to compromise the integrity of evidences. In this paper, the existing forensics techniques are reviewed including computer, mobile, network, clouds, digital, malware, and e-mail forensics, and the anti-forensics techniques and activities are also described and classified.
Recently, the anti-anti-forensics techniques appeared to defend the forensics tools and techniques against anti-forensics activities. Therefore, the importance of forensics and anti-anti-forensics will be highlighted to identify the anti-forensics attempts that target the evidence.

\par
Lately, not enough focus was given to the forensics domain, especially, in some fields such as the Internet of Things (IoT) and Cyber-Physical Systems (CPS). Therefore, this paper aims to develop a robust knowledge about recent forensics and anti-forensics approaches, methods, techniques, tools, towards reducing and preventing the causes and consequences of cyber attacks. Therefore, one of the paper objectives is to provide a better understanding of the digital forensics and anti-forensics domains.

\subsection{Contributions}
This paper contributions can be summarized in the following points:
\begin{itemize}
\item \textbf{Identifying \& Classifying:} digital forensics domains, techniques, tools, while also presenting their different approaches along their limitations.
\item \textbf{Identifying \& Classifying:} anti-forensics activities, techniques and tools, along their outcomes and consequences.
\item \textbf{Including:} the limitations and challenges that digital forensics investigators encounter during their investigation. 
% \item \textbf{Classifying:} cyber-threats according to their source and type.
\item \textbf{Discussing:} Counter Anti-Forensics (CAF) or Anti-Anti-Forensics (AAF) in terms of detection using machine learning methods, and in terms of prevention using privacy preserving techniques to protect digital evidences.
\item \textbf{Highlighting:} the most persistent digital forensics challenges.

\item \textbf{Proposing:} various suggestions and recommendations which are included to overcome the existing challenges and to enable efficient forensics investigations. 
\end{itemize}

\subsection{Related Work}
Many surveys were conducted and solely focused on cyber-crimes from a forensics viewpoints \cite{zhang2012survey}, while separately discussing digital forensics tools depending on the forensics type (network, malware, memory, etc.)~\cite{wazid2013hacktivism,pilli2010network,rogers2004future}. Anti-forensics techniques were presented in~\cite{gul2017survey,zhang2011survey,felt2011survey}. In~\cite{chen2014big}, big data challenges were presented as a serious forensics issue.
This paper presents a detailed analytical understanding of forensics (chain of custody, evidence source, forensics types, available tools and approaches) and anti-forensics (aspects, techniques, tools and approaches) domains, while being among the first to discuss the anti-anti-forensics aspect. Moreover, a broader range of forensics challenges is presented. Finally, privacy preserving of digital evidences' aspect was also presented in this work from a forensics viewpoint in respect to users privacy and evidences integrity. In other terms, this paper covers all the topics that are separately presented by other work, and also develops its own perspective on forensics/anti-forensics techniques. This will help fellow researchers and colleagues to broaden their search and understanding abilities. 

\subsection{Problem Formulation}
Cyber-crimes are expanding daily, and the use of anti-forensics techniques and activities is also increasing. As a result, this made it difficult to retrieve traces and gather evidences in regards of starting a forensics investigation. This paper aims to identify the different forensics and anti-forensics approaches for a better understanding and further enhancements and precautions against anti-forensics activities.

\subsection{Organization}
%TO BE ADDED> 
This paper is divided into x sections aside the introduction and is presented as follows:
%In section~\ref{sec:1}, a forensics background is presented including digital data, forensics investigators, cyber-crimes and cyber-threats classification.
%In section~\ref{sec:2}, different forensics sub-domains are presented and discussed according to their available approaches and techniques in use.
%In section~\ref{sec:3}, digital forensics challenges are highlighted and discussed.
%In section~\ref{sec:4}, anti-forensics aspects and techniques are presented and discussed.
%In section~\ref{sec:5}, anti-anti-forensics or counter anti-forensics detection and prevention techniques are presented and discussed.
%In section~\ref{sec:6}, various suggestions and recommendations are proposed.
%In section~\ref{sec:7}, this paper's work is concluded with an insight on the future work.

\section{Background - Forensics Domain }
~\label{sec:1}

In this section, the forensics data classification is highlighted, while the cyber-crimes aspects is discussed from a forensics point of view, in addition to identifying and classifying cyber-threats.

\subsection{Forensics Data Classification}
 Till now, there is no unique uniform standard that classifies forensics data. However, some of the existing classifications are somewhat similar. In \cite{halboob2015privacy}, Halboob et al. identified the forensics data as being directly accessible data (DAD), privacy-preserved accessible data (PAD), or non-accessible data (NAD). However, in a more granular way, forensics data can be classified into four main categories (see \figurename~\ref{fig:e}), as listed in the following: 
 
\begin{figure*}[!htp]
 \centering
    \includegraphics[scale=0.75]{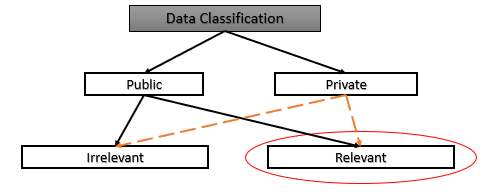}
    \caption{Forensics Data Classification}
    \label{fig:e}
\end{figure*} 
\begin{itemize}

\item \textbf{Public \& Irrelevant Data:}
This type of data does not present any useful information. It aims to ensure that investigations are more time consuming and less accurate.
\item \textbf{Public \& Relevant Data:} 
This data can be modified and wiped beyond recovery by cyber-criminals to evade any detection and eliminate any data that can serve as a possible evidence.
\item \textbf{Private \& Irrelevant Data:}
This data can be encrypted or hidden by using data hiding techniques (E.g steganography). 
\item \textbf{Private \& Relevant Data:}
 This type of data serves as a hidden treasure for forensics investigators. In fact, this data can reveal many information about the attackers, including source, attack fingerprints, skills, experience and strategies, which can help track and identify them. 
\end{itemize}

\subsection{Forensics Investigators:}
Depending on the source of the retrieved data, forensics investigators can be classified into four categories: 
\begin{itemize}
\item \textbf{Logical Forensics Investigators:} or Digital Forensics Investigators (DFI) are concerned  about retrieving evidences from digital devices, including software, operating systems, Portable Computers (PCs), laptops, or even smart-phones found at a crime scene.
\item \textbf{Cyber Forensics Investigators:} are involved in the world of IoT and its different fields/domains, including cloud services.
\item \textbf{Physical Forensics Investigators:} or traditional investigators rely on their expertise, knowledge, and experience to retrieve physical forensics evidences, mainly from hardware equipment and devices.
\end{itemize}

\subsection{Cyber-Crimes}
It is essential to identify a committed cyber-crime's life-cycle to achieve a criminal goal. 
In the following, the cyber-crimes, cyber-threats and cyber-attacks are discussed in details. 

\subsubsection{Cyber-Crime Steps}
\begin{figure*}[!ht]
  \centering

    \includegraphics[scale=0.7]{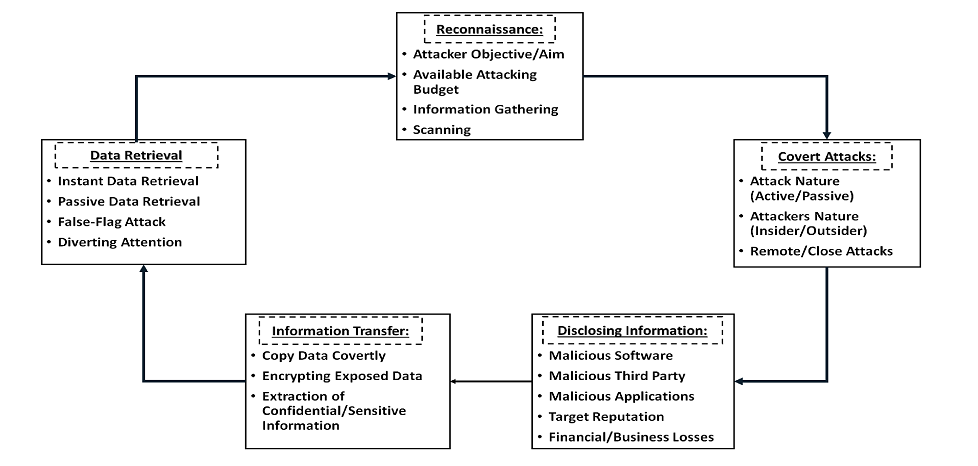}
    \caption{Cyber-Crime Cycle}
    \label{fig:cc}
\end{figure*}

Cyber-crimes are committed using predefined steps to commit various crimes such as hacking, phishing, spamming, or to commit cyber-offense hate crimes, blackmails, bullying child/adult pornography \cite{shariff2008cyber,finkelhor2004child,johnson2016cyberbullying,smith2008cyberbullying,donegan2012bullying}. This paper presents the cyber-crime's life-cycle as follows (see \figurename~\ref{fig:cc}), which is a modified version of the main cyber-crime phases presented in~\cite{stockdale2004benefits}.

\begin{itemize}

\item \textbf{Reconnaissance:}
This phase consists of collecting information by the attacker about the targeted victim(s) (individuals or organizations) according to an attacker's objective, available budget and resources.
\item \textbf{Covert Attacks:}
Once the information is gathered, attacking plans and strategies are prepared according to their available tools and techniques. Most of these attacks are conducted through covert channels using Virtual Private Networks (VPNs), The Onion Router (TOR) or proxies. 
These attacks can be performed by:
\begin{itemize}
\item \textbf{Insiders:} or whistle-blowers, by being a rogue or unsatisfied employee recruited by a competitive organization to target a rival organization. 
\item \textbf{Outsiders:} by leading a covert remote attack through spamming emails that lure an employee to click on malicious links, or visit malicious websites. 
\end{itemize}
Finally, the types of attacks that can be performed by an attacker, in this phase, can be divided into two types: 
\begin{itemize}
\item \textbf{Direct Human Interaction Attacks:} are performed through social and reverse engineering attack types~\cite{krombholz2015advanced,lee2011tie}.
\item \textbf{Indirect Human Interaction Attacks:} are performed through phishing, spear phishing, whaling and vishing~\cite{badra2007phishing}. 
\end{itemize}

\item \textbf{Disclosing Information:}
Data/information disclosure is possible by using a covert malicious software/application being installed on the victim’s system either through spamming or surveillance attacks to reveal business trade secrets, target an organization’s reputation, intellectual property theft, or causing huge financial losses.

\item \textbf{Information Transfer:} 
This phase includes copying the exposed information to a place, where it is easier for the attacker to manipulate it without being detected. Most of the time, the exposed data might be encrypted, where decrypting would be time consuming.

\item \textbf{Data Retrieval:} 
Data can be retrieved in three main ways~\cite{stockdale2004benefits}:
\begin{itemize}
\item \textbf{The first way:} is based on retrieving data instantaneously, but with a higher risk of being detected.
\item \textbf{The second way:} is based on retrieving data passively, which is a time consuming process but without the risk of being detected. 
\item \textbf{The third way:} is based on creating a false-flag attack to divert the attention of the attacker towards another incident.
\end{itemize}
This bides attackers more time to retrieve the needed data and cover their tracks, with organisations taking months and even years to recover from their losses. 
\end{itemize}

\subsubsection{Cyber-Criminal Classification}
Cyber-criminals can be classified into three main categories: 
\begin{itemize}
\item \textbf{Worldwide Organized Groups:}
are geographically separated groups that operate (jointly/separately) locally, regionally, or even globally to cause a bigger damage against their intended target(s).
\item \textbf{Regional Organized Groups:}
 operate locally or regionally within a limited geographical area and are limited in terms of experience, knowledge, skills and available tools (script kiddies). 
\item \textbf{Individuals:}
are known as lone wolves and mainly operate locally. Their ability is limited especially in terms of available manpower and skills to perform coordinated tasks.

\end{itemize}

\subsubsection{Cyber-Crime Structure}
A cyber-crime's type is defined depending on the cyber-criminal’s skills, knowledge, experience, available tools, resources, and manpower. This also includes the level of structure, communication, collaboration and cooperation between different cyber-criminals. In this regard, this paper classifies cyber-crimes into four main types.
\begin{itemize}

\item \textbf{Coordinated \& Organized:}
Cyber-crimes are coordinated if they are performed simultaneously and in a professional, experienced and synchronised manner. These crimes are mainly based on Web defacement or series Distributed Denial of Service (DDoS) attacks compromising the availability of a given organization and preventing legitimate users from gaining any access for a given period of time. This causes serious financial and economic losses. These cyber-crimes can also be used to perform a cyber-heist against banks~\cite{shevchenko2016cyber,simonovski2018financial}.\\
Organized cyber-crimes are conducted to exploit organizations' vulnerabilities and security gaps to ensure a high profit at a lower risk~\cite{general2013commonwealth} through the coordination and the collaboration between well-trained and skilled cyber-criminals. This is done using anonymous covert connection types such as The Onion Router (TOR)~\cite{perry2013design} and the deep dark web~\cite{chen2008uncovering,greenberg2014hacker} to evade any detection.

\item \textbf{Uncoordinated \& Disorganized:}
Uncoordinated crimes can be easily detected due to flaws in the carried out attacks. This is due to the lack of experience, knowledge, communication, and synchronization between cyber-criminals. As for disorganised attacks, there is a possibility to loose the attack’s tracks and fail to eliminate any possible evidence source. Thus, unfulfilling the attacker's main objective.

\item \textbf{Conventional:} These crimes can be predictable, due to following a certain attack or hacking cycle, which makes it easier to identify them. In fact, they can have serious impact and implications whenever they are conducted, especially if they are coordinated and organised.

\item \textbf{Unconventional:} This class of cyber-crimes is unpredictable and can be divided into two main types. The first type is related towards conducting highly advanced attacks through the exploitation of unknown security gaps. The second type is based on using advanced anti-forensics tools to erase data beyond recognition and recovery.
\end{itemize}

\begin{figure*}[!htp]
  \centering
    \includegraphics[scale=0.3]{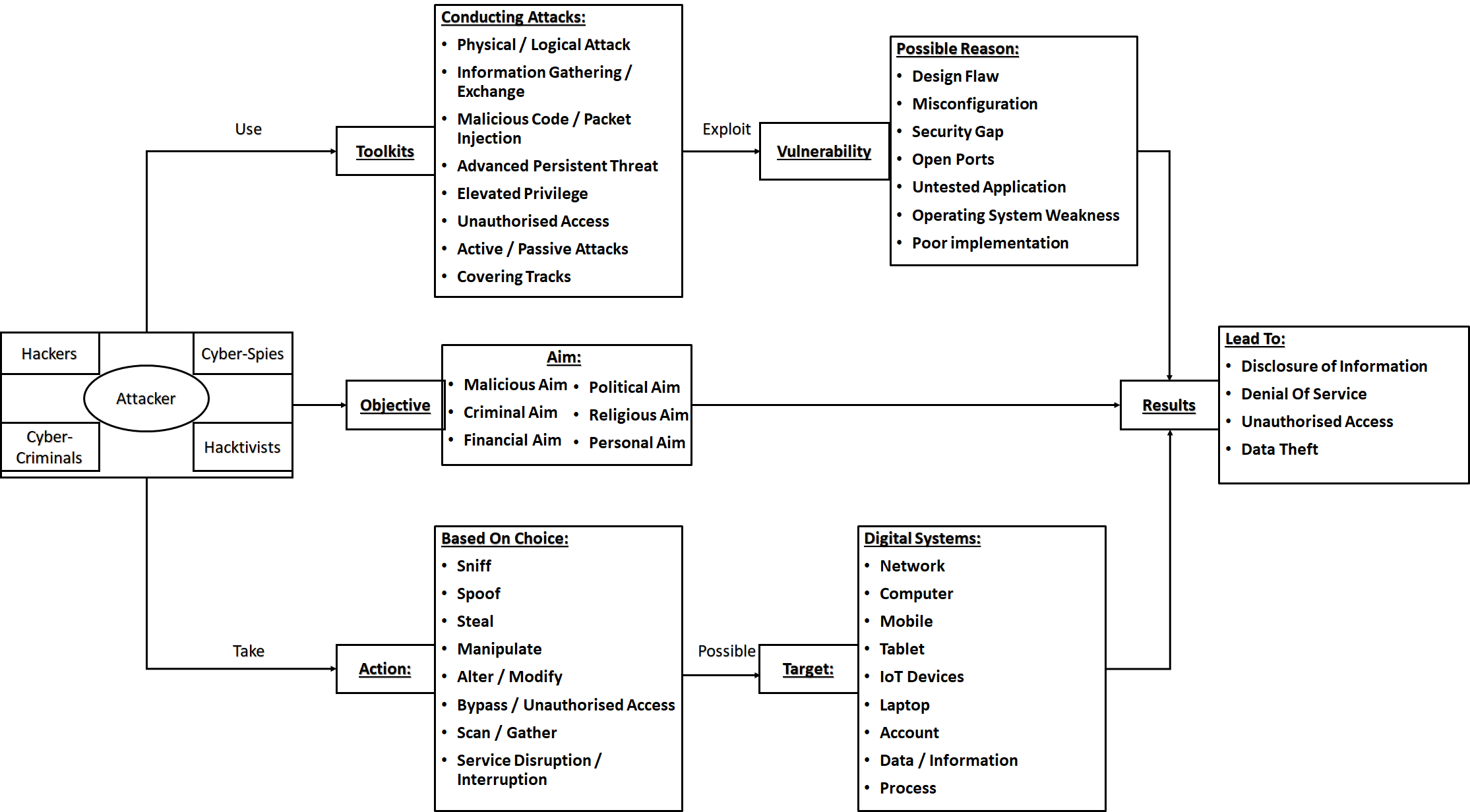}
    \caption{Classification of Cyber Attackers}
    \label{fig:q3}
  
\end{figure*}

\figurename~\ref{fig:q3} includes the attackers' types according to their intentions.

\subsection{Cyber-Threats}

\begin{figure*}[!ht]
  \centering
  \begin{minipage}[b]{0.85\textwidth}
    \includegraphics[width=1\textwidth]{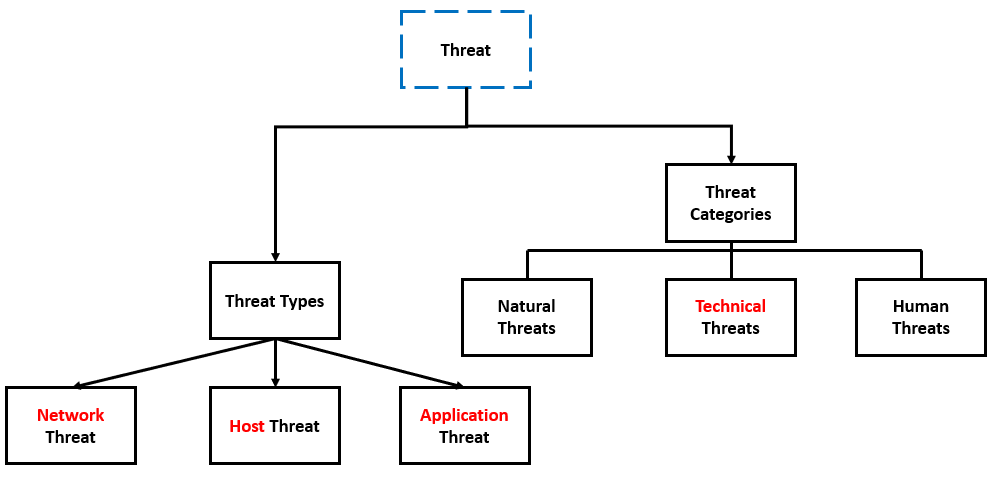}
    \caption{Threat Identification \& Classification}
    \label{fig:ThreatIdentification}
  \end{minipage}
\end{figure*}

Threats occur from the risk and potential of having the occurrence of an accident, or a given attack. In fact, the types of the existing threats can be summarized in \figurename~\ref{fig:ThreatIdentification}. However, it is still not enough to really understand the source and nature of the attack.

\subsubsection{Threat Source}
A cyber-threat source cannot be always easily tracked due to criminals relying on anonymous ways to perform their attack. However, it can be classified as one of the following categories:
\begin{itemize}
    \item \textbf{Cyber-Criminals:} are usually organised hacking groups or individuals that conduct cyber-heist, bullying, blackmailing or leaking private (financial, military, medical or governmental) information to malicious third parties through the deep dark web for personal or monetary gains.
    \item \textbf{Hacktivists:} are usually hackers that aim to launch (distribute) denial of service or web defacement attacks as part of creating a cyber-protest against a political party or a government.
    \item \textbf{Cyber-Terrorists:} aim to perform web-defacement and information leakage attacks targeting organisations, oil industries, governmental and military installations.
    \item \textbf{Cyber-Spies:} usually target organisations, enterprises, governmental and military installations as part of conducting espionage or/and sabotage operations.
\end{itemize}

\subsubsection{Threat Type}
Depending on their nature, threats can be identified differently. For example, Stoneburner et al. defined a threat as being a potential security breach that aims to accidentally trigger or exploit a specific vulnerability or security gap~\cite{stoneburner2002sp}. It is somehow similar to a technical exposure. Therefore, identifying threats is not a straightforward task. However, this paper presents them as follows:
\begin{itemize}
    \item \textbf{Security Breach:} is usually caused by the presence of weak security measures (gaps), or limited security measures that only cover parts of the exploited system and not apply the defense-in-depth mechanism, unless its a zero-day attack attempt.
    \item \textbf{Cyber Attack:} is primary caused through the exploitation of a given system, device or/and information via wireless communications including networks and internet, by targeting its confidentiality, integrity, availability or/and authenticity.
    \item \textbf{Physical Attack:} is caused by criminals masqueraded as employees that breach into a given organisation and physically damage its systems and devices.
    \item \textbf{Untested Application:} applications that are untested are prone to various malicious and non-malicious security breaches, including malfunctioning, abnormal performance, backdoors, rootkits and malwares including viruses and Trojans. 
    \item \textbf{Old Version Systems:} that are not constantly/regularly updated are usually targeted by already known exploits/attacks, leading to various security/privacy breaches.
    \item \textbf{Exploitable Vulnerability:} is an exploitable gap found in a given security program which upon its exploitation it allows a given attacker to gain an unauthorized access to a given system/device.
\end{itemize}

\subsection{Forensics Chain of Custody}

To accomplish a forensics investigation, a Chain-of-Custody~\cite{hargreaves2012automated} should be followed by any forensics investigator. This chain consists of four phases, and they are described in the following: 
% as shown in \figurename~\ref{fig:forensics Investigation Phases}.
% \begin{figure*}[!ht]
%   \centering
%   \begin{minipage}[b]{0.5\textwidth}
%     \includegraphics[width=\textwidth]{6}
%     \caption{Digital Forensics Investigation Phases}
%     \label{fig:forensics Investigation Phases}
%   \end{minipage}
% \end{figure*} 

\begin{itemize}
\item \textbf{Identification:} includes identifying the event and identifying the evidence. To investigate an incident, two types of investigation skills are needed: \textbf{soft investigations skills} and \textbf{hard investigation skills}. This will achieve a successful investigation to identify what happened, where it happened, how it happened, who was targeted, and who was the attacker. These two types of skills can be defined as follows:

\begin{itemize}
\item \textbf{Soft Digital Forensics Investigative Skills (SDFIK):} include a strong cooperation and collaboration between forensics investigators, the public, and the investigation team. Its main task is to assess a situation promptly and to help, identify and differentiate between normal and suspicious events. 
\item \textbf{Hard Digital Forensics Investigative Skills (HDFIK):} require the collection of information from any public domain, while maintaining the needed awareness level, following the chain-of-custody.
%which is achieved through a high level of strictness and compliance.
\end{itemize}

\item \textbf{Data Collection:} can be done, either by communication monitoring, or through communication interception.
In this phase, all data types are collected to identify any potential evidence. The original data must first be copied, and then, all the forensics work can be performed on the copied data once the hashes are compared and matched. 
\begin{itemize}
\item \textbf{Monitoring Communications:} 
is usually achieved by conducting an “intrusive surveillance”~\cite{donner1980age,maguire2000policing} (i.e covert vehicle~\cite{ross2007place}, mainly a covert spying van), conducting a “directed surveillance”~\cite{fernandez2011determining} (i.e relying on smart street or security cameras), or using human agents~\cite{bernal2016data}.
\item \textbf{Intercepting Communications:} is usually achieved by identifying and intercepting IP/MAC addresses of cyber-criminals and suspects alike \cite{gorge2007lawful,kipper2007wireless}, in addition to tracking their e-mails/web-activities, and identifying their phone numbers and monitoring their phones messages, calls and logs (i.e land-lines and smart-phones~\cite{bryant2016policing}). 
\end{itemize}

\item \textbf{Analysis \& Evaluation:}
Upon completion of initial investigation phases, it is also important to analyze what happened in order to evaluate the type of crimes committed.
This phase consists of two sub-phases which are explained as follows:  
\begin{itemize}
\item \textbf{Evaluation \& Analysis:} 
This phase consists of analyzing the retrieved data from software/hardware equipment including Hard or Solid State Drives (SSD), virtual machines, networks, and network devices, smart-phones, tablets, and laptops \cite{giova2011improving}. This also includes conducting a SIM card analysis based on identifying contact lists, call logs, and Short Message Services (SMS) (mainly recent, or around a given crime event), before classifying it as intellectual property theft, stalking or/and threatening behaviour, cyber-sexual offences, unauthorized access, sabotage, espionage, and/or disclosure of sensitive information, etc...

\end{itemize}

\item \textbf{Reporting:}
 is the last phase of a forensics investigation. Thus, it is divided into two main steps:
 \begin{itemize}
\item  The first step consists of the evidences collection to follow a \textbf{legal process of prosecution} to prove that the suspected criminal is guilty.
\item The second step is the \textbf{legal process to prove the evidence's legality}. A legal evidence is based on identifying whether a given fact can be proved and backed or not. This is done by checking if the evidence is supported by real facts to prosecute a given suspect. In some cases, the evidence can be returned~\cite{daniels2008forensic}.
  \end{itemize}

\end{itemize}

\par
In the next section, the paper dives deeper in the digital forensics domain, reviewing its different sub-domains.

\section{Digital Forensics Sub-Domains}
~\label{sec:2}
\begin{figure*}[!ht]
   \centering
   \begin{minipage}[b]{1\textwidth}
     \includegraphics[width=\textwidth]{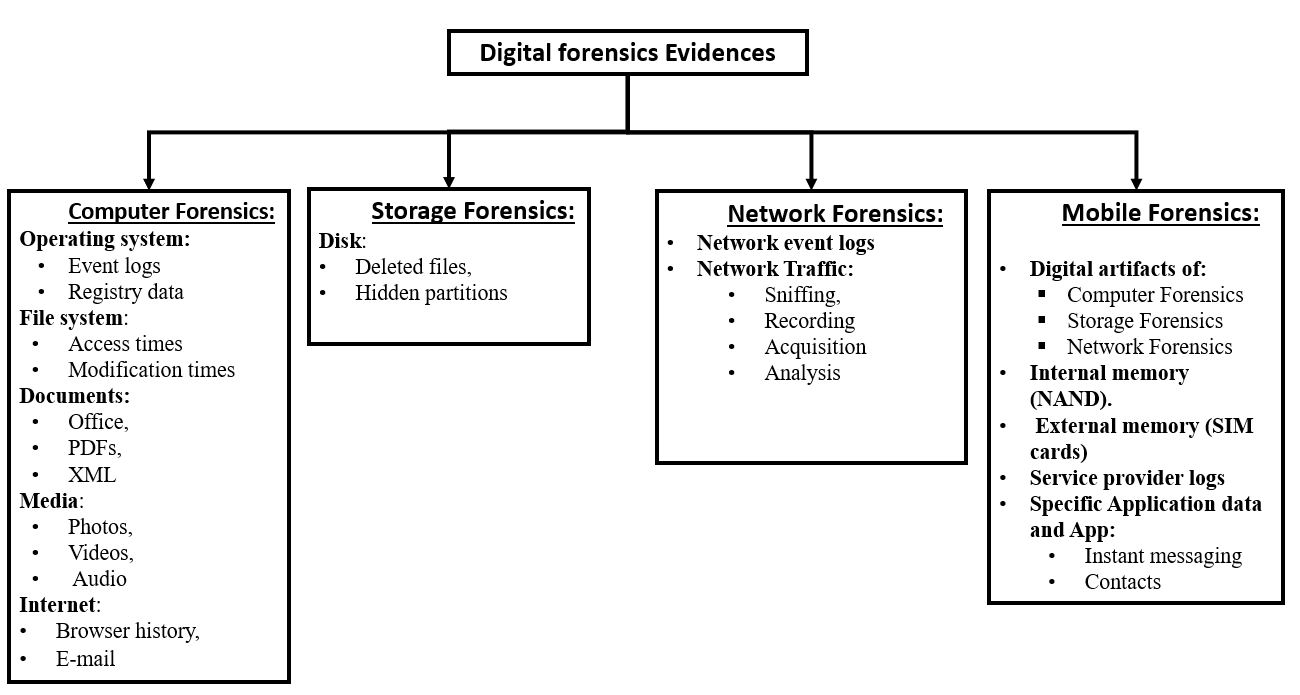}
     \caption{Classification of Digital Forensics Artifacts}
     \label{fig:DFs}
   \end{minipage}
 \end{figure*}

\begin{figure*}[!htp]
  \centering
  %\begin{minipage}[b]{0.8\textwidth}
    \includegraphics[scale=0.5]{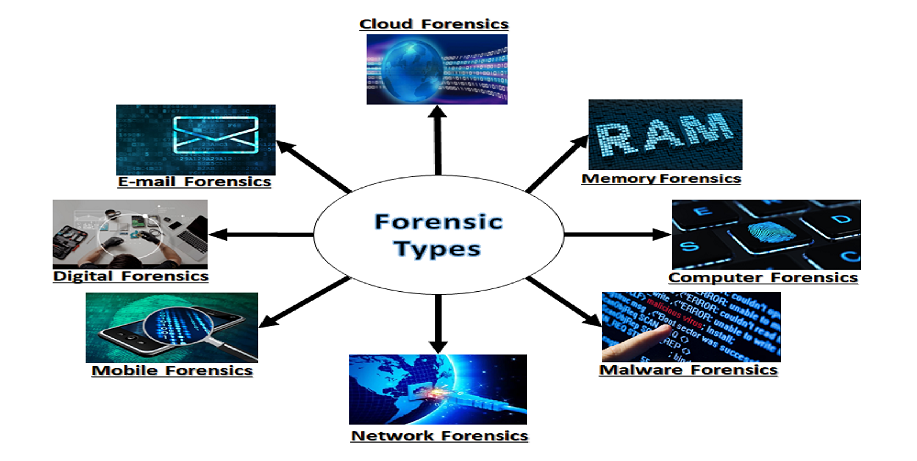}
    \caption{Forensics Types}
    \label{fig:q1}
 % \end{minipage}
\end{figure*}

Digital forensics are used to uncover and interpret electronic data related to a cyber-crime. The aim is to preserve the evidence to be legally used in courts without any alteration/modification. In fact, digital forensics evidences can be retrieved from various digital forensics sources, as illustrated in \figurename~\ref{fig:DFs}. Additionally, there are several forensics types, as presented in \figurename~\ref{fig:q1}, which will be discussed in the following sections.
 
%\par
%Therefore, digital forensics can be divided into 8 sub-domains and they are presented in the following. the detection of the attack related traces.

\subsection{Computer Forensics}

Cyber-crimes will lead to more than \$2 trillion losses in business by 2019~\cite{benitez2018computer}. In fact, more than 4,762,376,960 data records were either lost or stolen since 2013. As a result, the demands for Computer Forensics (CF) techniques and tools arose respectively.
Through investigations, computer forensics collect data from computer-based devices.
This allows them to check the running processes on the attacker/victim computer through system file checking and extraction, which makes it possible to track and trace a given attack. Therefore, CFs read the hard disk information derived from digital sources within a short time span~\cite{rogers2006computer}, before any deletion/modification occurs. %of the attack traces. 

%. In addition, In fact, . computer forensics. can be.

\subsubsection{Available Tools}
Since computer forensics help with identifying and tracking down attackers~\cite{wazid2013hacktivism}, different tools started being used (see \tablename~\ref{tab:1}), including FTK imager, Encase, Helix, Winhex, SMART, X-Ways and Autopsy, along with other tools listed in~\cite{kambalapalli2018different,casey2011digital,tabona2002top,rogers2004future}. However, computer forensics still present various challenges, limitations and drawbacks that need to be considered~\cite{bennett2012challenges,schweitzer2003incident,yasinsac2003computer}.
\begin{table*}[!ht]
\centering
\scriptsize
\caption{\textbf{Computer Forensics Analysis Tools}}
\label{tab:1}
\begin{tabular}{|p{3.5 cm}|p{12.5cm}|}
\hline
\centering
\textbf{Computer Forensics Analysis Tools (CFATs)} & \textbf{Description} \\ \hline
Autopsy & A GUI-based open source digital forensics program that analyses hard drives \& smartphones \\ \hline
Encrypted disk detector &  It serves to to check encrypted physical drives \\ \hline
Magnet RAM capture &  It is used to capture a computer’s physical memory and analyses memory artefacts\\ \hline
RAM capturer & It is used to dump data from a computer’s volatile memory which may contain log in credentials \& Encrypted password volumes \\ \hline
Splunk & Combines many tools including WHOIS/GeoIP lookup along pinging and port scanning along other tools \\ \hline
Forensics Acquisition of Websites (FAW) & It captures the entire or partial page, along HTML source code and all images type  \\ \hline
USB write blocker & It extracts the USB drives content without leaving any fingerprint  \\ \hline
NFI Defraser & It detects partial or/and full multimedia files in data streams \\ \hline
ExifTool & It Helps investigators to read, write and edit meta information for various file types \\ \hline
Toolsley & File identification and signature verification, along Data URI and password generating, and other useful investigative tools \\ \hline
SANS Investigative Forensics Toolkit (SIFT) & Among the most popular open source incident response platforms \\ \hline
Dumpzilla & It extracts different browser information and analyses them \\ \hline
Foxton & It can be a browser history capturer or a browser history viewer\\ \hline
ForensicUserInfo & Extracts NT Hash, login counts, profile paths account expiry date, password reset and other infro \\ \hline
Black Track & It is used as a pen testing and a forensics tool \\ \hline
Paladin & Has a variety of forensics tools needed to investigate any incident \\ \hline
Sleuth Kit & Collection of command line  tools  \& It consists of analysing volumes \& files \\ \hline
 Computer Aided Investigate Environment (CAINE) & It is used to analyse, investigate and create a forensics report using a variety of tools \\ \hline
Encase & Can access a large file system number of file system whilst creating timestamps \\ \hline 
Forensics Tool Kit (FTK) & It analyses different file systems whilst revealing their different timestamps \\ \hline 
Zeitline & It is known as timeline editor, collects evidences from log files to solve digital crime cases \\ \hline
Cyber Forensics Timelab (CFT) & It scans \& views hard-drives whilst identifying the sorted timestamp of the found file     \\ \hline
Crowd Response & It gathers a system’s information to initiate an incident response along with security engagements~\cite{boyes2014trustworthy} \\
\hline
Registry Recon & It extracts registries information (previous/current) from the evidence \& rebuilds their representation \\ \hline
Llibforensics & It is a library for digital forensics applications, developed in Python, extracts information from different evidence types \\ \hline
Coroner’s Toolkit & It is used to aid analysis of computer disasters and data recovery \\ \hline
Computer Online Forensics Evidence Extractor (COFEE) & It gathers evidence from Windows systems, and can be installed on a USB pen drive or external hard disk with 150 different tools \& a GUI based interface for command \\ \hline
HELIX3 & It is used in incident response, made up of many open source digital forensics tools including hex editors \& password cracking tools \\ \hline
PlainSight & It includes viewing Internet histories, checking USB device usage, extracting password hashes \& Information gathering along other tasks, like analysing data collected from physical memories \\ \hline

\end{tabular}
\end{table*}
%the integrity of the data for an investigation process. . This is done by relying on various methods and techniques. These methods might include a software analysis to search through data archives. , downloaded or even uploaded. 
\subsubsection{Computer Forensics Steps}
 Several computer forensics steps must be considered to ensure a successful investigation. In~\cite{kumari2016insight}, Kumari et al. discussed some computer forensics steps like sanitizing the storage area to protect forensics processes, and the use of Message-Digest algorithm (MD5 hash) image’s calculated value to find out whether the image used is original or not. As a result, computer forensics can be divided into five main steps, also mentioned in~\cite{beebe2005hierarchical} including:
\begin{itemize}
\item \textbf{Developing Policies \& Procedures:} digital evidences can be complex and sensitive, where data can be easily compromised if not carefully handled and protected. Therefore, establishing standard policies and procedures with strict guidelines can help support and enhance a computer forensics investigation.
\item \textbf{Assessing Evidences:} evidence/potential evidence assessment reveals a clear understanding of the committed cyber-crime details. Therefore, new methods needs to be adopted to assess any potential information serving as an evidence, including the digital evidence type and its format.
\item \textbf{Acquiring Evidences:} requires a very detailed plan to acquire data legally. However, documenting data is recommended before, during and after any acquisition process. This allows essential information (based on software and hardware specifications) to be recorded and preserved by maintaining data's integrity.
\item \textbf{Examining Evidences:} includes examining the data that is copied, retrieved and stored in databases from a given designated archive, by relying on specific key words or/and file types/names even if these files were recently deleted. This offers the chance to know when the data was created or/and modified.
\item \textbf{Documenting \& Reporting:} 
An accurate record of the activities is kept by computer forensics investigators. This includes methods of retrieving, copying and storing data, along with acquiring, examining and assessing the evidence. Therefore, preserving the data integrity, enforcing the right policies and procedures, and authenticating any findings (how/when/where) related to the recovered evidence.

%. Each event is stored in a database table along with its related information. In fact, CAT Detect . %However, any manipulation of time can seriously disable the forensics investigation.  . In fact, such tools help investigators to know the events or evidences to facilitate the forensics investigation. This.However, cyber-criminals can also use the Timestap tool to clear out all the file system timestamp history.

\item\textbf{Computer Activity Timeline Detection:}
Computer Activity Timeline Detection (CAT Detect)~\cite{marrington2011cat,al2013cat} is based on analysing computer activities to detect inconsistencies in a computer system timeline. However, the investigation process is prone to data, event, or even file loss either through deletion, manipulation or overwriting. "CAT Detect" can remove inconsistencies in a given timeline by parsing the window system event logs, and can access the MAC (Modified Accessed Created) file metadata, and create a database table related to the information file being accessed. Thus, making it possible to build the evidence through time-line construction.
\item\textbf{Computer Forensics Timeline Visualisation:}
Computer Forensics Timeline Visualisation (CFT Visual) is a timeline based tool used in the Cyber-forensic Timelab~\cite{olsson2009computer}. By applying this method, the obtained evidences are based on the time variations which would result into creating a timeline based graph of events that allows the investigators to know and identify what happened right after and before a given event (cause/consequence). 
\end{itemize}

%\subsection{Anti-Forensics}

%\subsection{Available Solutions}

%\subsection{Recommendation}

\subsection{Network Forensics}
Network-forensics is a branch that derives from the digital forensics tree. It is responsible of monitoring and analysing devices and network traffic as part of legal information gathering and legal evidence retrieval. %Unlike other forensics branches, network forensics operate with volatile and dynamic information. 

\subsubsection{Network-Based Evidence}
\begin{figure*}[!ht]
  \centering
    \includegraphics[scale=0.5]{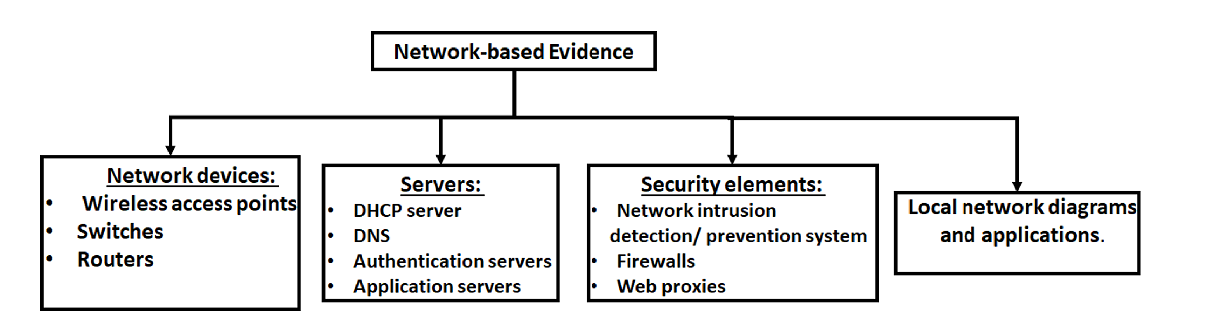}
    \caption{Network Based Evidence}
    \label{fig:nb}
\end{figure*}

Network forensics tools are used to retrieve network-based evidences, which are categorized as follows (see \figurename~\ref{fig:nb}).
\begin{itemize}
    \item \textbf{Network Devices:} include wireless access points (APs), switches and routers, with their IP and MAC addresses pools.
    \item \textbf{Servers:} include DHCP/DNS Servers, as well as authentication and application servers, with the usernames, passwords, activities, and privileges.
    \item \textbf{Security Elements:} include Web proxies, network Intrusion Detection/Prevention Systems (IDS/IPS), as well as stateful and stateless firewalls to monitor incoming and outgoing internet traffic and saved logs.
    \item \textbf{Local Networks:} include diagrams and used applications, with their logs that contain details and information about the connected users.
\end{itemize}

\subsubsection{Network Forensics Approaches}
 Due to the increase in the network misuse behaviour, new approaches were needed to ensure a quick emergency response and incident investigation, to enhance the network security and forensics aspects. As a result, an evidence graph was presented by Wei et al. in~\cite{wei2005network}. In~\cite{kumari2016insight}, Kumari et al. reviewed different approaches used for network forensics investigations. In~\cite{alzaabi2015cisri}, Alzaabin et al. presented a forensics analysis system used for crime investigation. The authors relied on the "Crime Investigation System Using the Relative Importance of Information Spreaders in Networks Depicting Criminals Communications" (CISRI) framework~\cite{alzaabi2015cisri} to determine the head of the cyber-criminal group through crime investigation and to describe the relationship between a given criminal group through a graph that is based on phone calls, call logs and messages. In~\cite{hajdarevic2015approach}, Hajdarevic et al. presented a different approach that allows the collection, the analysis and the reporting of evidences through a forensics investigation. This helps making the digital evidence legal and acceptable by courts. In~\cite{bijalwan2015forensics}, new network forensics prototype tools were presented to generate an automated evidence analysis. This helps reducing the response time to overcome the issues of manual analysis. In fact, these prototypes would help identify the attackers and determine each one’s role in a given cyber-criminal group. In \cite{koroniotis2019forensics}, Koroniotis et al. managed to provide a new network forensics taxonomy that is applicable to botnets in IoT-related and non-IoT-related domains to assess their strengths and weaknesses. In~\cite{pilli2010network}, Pilli et al. developed a network forensics analysis framework known as “Network Forensics System” including forensics analysis, and network security and monitoring tools~\cite{pilli2010network,hunt2012network}, to seize cyber-criminals in cyber-space~\cite{davidoff2012network}. 

\par
%. According to. , network forensics were categorized into two main categories.
In fact, network forensics can be categorized differently, or mainly as two categories explained in~\cite{yurcik2003two,mukkamala2003identifying}. In the first category, \textbf{“Catch It As You Can (CIAYC)”}~\cite{garfinkel2002network}, all packets are sent through a traffic point before being stored into databases for further analysis. In the second category, \textbf{“Stop Look And Listen (SLAL)”}~\cite{garfinkel2002web}, data is only stored in databases for future analysis.

\subsubsection{NFA \& NSM Tools}
Network Forensics Analysis Tools (NFATs)~\cite{corey2002network} and Network Security and Monitoring Tools (NSMTs)~\cite{yurcik2003two} are used to analyse the collected and aggregated data. Moreover, they provide IP security, along with the detection of insider/outsider attacks, as well as ensuring a risk analysis, and data recovery. This allow them being used for anomaly detection with the ability to detect and predict future attacks through the reliance on IDS/IPS and Firewall logs~\cite{shrivastava2016network}. Therefore, the main network forensics tools are summarized in \tablename~\ref{tab:2} as follows:
\begin{table*}[!ht]
\scriptsize
\centering
\small
\caption{\textbf{ Forensics tools used for network forensics investigations}}
\label{tab:2}
\begin{tabular}{|p{3.5cm}|p{12.5cm}|}
\hline
\centering
\textbf{Network Forensics Analysis Tools (NFATs)} & \textbf{Description} \\ \hline
 NetIntercept & It captures network traffic, stores it in a pcap format, detects spoofing, and generates a variety of reports.
 \\ \hline
 NetWitness & It captures network traffic and reconstructs the network sessions to the application layer to ensure an automated analysis and zero-day detection. \\ \hline
 NetDetector & It captures intrusions and performs multi time-scale network analysis and signature-based anomaly detection. \\ \hline
 Iris & It collects network traffic and reassembles it, also it has an advanced search and filtering mechanism for a quick data identification.
 \\ \hline
 Infinistream & It utilizes intelligent Deep Packet Capture (iDPC) technology, whilst performing a real-time or back-in-time analysis, and smart recording and data mining for optimization. 
  \\ \hline
 Solera DS 5150 & It is used for a high-speed data capture, and network traffic filtering
 \\ \hline
 DeepSee & It includes three softwares which are reports, sonar and search, to index, search and reconstruct network traffic.
  \\ \hline
 OmniPeek/Etherpeek & It provides a real-time network visibility, along with a high capture capabilities and expert analysis, also it ensures a low-level network traffic analyser. \\ \hline
 SilentRunner & It captures, analyzes and visualizes network activity, whilst reconstructing security incidents in their exact sequence. \\ \hline
 NetworkMiner & It captures network traffic through real-time or passive sniffing, and assesses how much data was leaked.\\ \hline
 Xplico  & It captures Internet traffic and reconstructs it to present the results in a visualized form.\\ \hline
 PyFlag & It analyses network captured packets, whilst also supporting a number of network protocols, along parsing pcap files, to extract and dissect packets at low level protocols. \\ \hline
TCPDump & I is a windump command-line network packet analyser supporting network forensics analysis.\\  \hline
Ngrep & It is a tool that debugs a low-level network traffic. \\ \hline
Wireshark &  It forms a basis of network forensics monitoring and forensics studies. \\ \hline
Airxxx-ng series & It is used to ensure a low-level traffic analysis tools for wireless LANs.\\ \hline
DeepNines & It provides a real-time identity-based network defense with basic network forensics.
 \\ \hline
Argus & It is used for non-repudiation purposes, along with the detection of slow scans, whilst supporting zero-day attacks.
\\ \hline
Fenris & It is used for code/protocol analysis, debugging, vulnerability research, security audits, network forensics and reverse engineering. \\ \hline
Forensics \& Log Analysis & It is used to ensure a log file analysis combined with network forensics, and Python implementation. \\ \hline
Snort & It is used for network intrusion detection and prevention, and network forensics analysis.
 \\ \hline
Savant & It is used for a live forensics/network analysis, along with critical infrastructure reporting. \\ \hline
Dragon IDS & It offers network/host intrusion detection, and forensics network analysis.\\ \hline
RSA EnVision & It provides a live network forensics analysis, data leakage protection and log management.\\ \hline
Solera DS & t provides a live network forensics analysis, used as an analysis tool along with the ability to extract metadata.\\ \hline
SLEUTH KIT &  It is a tool used to examine file systems and to detect suspicious computer behaviour without interrupting the network. \\ \hline

\end{tabular}
\end{table*}

A Further description of network security and monitoring tools is summarized in \tablename~\ref{tab:3}.
\begin{table*}[!ht]
\scriptsize
\centering
\small
\caption{\textbf{ Description Of Network Security \& Monitoring Tools}}
\label{tab:3}
\begin{tabular}{|p{2.5cm}|p{14cm}|}
\hline
\centering
\textbf{NSM Tool Names} & \textbf{Description} \\ \hline
TCPDump & A packet sniffer and analyser that runs in a command line, and intercepts and displays network transmitted packets. \\ \hline
Wireshark & A cross-platform tool that performs a live capture in libpcap format, along with an offline analysis. \\ \hline
TCPFlow &  A tool that captures transmitted data as part of TCP connections and stores it for a protocol analysis. \\ \hline
Flow-tools & A library that collects, sends, processes and generates NetFlow data reports. \\ \hline
NfDump &  It Works with NetFlow formats by capturing daemon reads, displays them and creates statistics of flows and stores the filtered data. \\ \hline
PADS &  A lightweight and intelligent network sniffer, and a signature-based detection engine used to passively detect any network asset.  \\ \hline
Nessus & A vulnerability scanner thar ensures a high-speed and sensitive data discovery and vulnerability analysis. \\ \hline
Sebek & Designed to capture all Honeypot activity \\ \hline
TCPTrace & It Produces different output types containing information, such as elapsed time and throughput. \\ \hline
Ntop & It is used for network traffic measurement, monitoring, planning, and detection of any security violations. \\ \hline
TCPStat & It reports the network’s bandwidth, along with the number of packets and the average packet size.\\ \hline
IOS NetFlow & It Collects IP attributes of each forwarded packet, and detects network anomalies and security vulnerabilities. \\ \hline
TCPDstat & It produces a per-protocol traffic breakdown, including network packets and traffic patterns.\\ \hline
Ngrep & It debugs plaintext protocol interactions to analyse and identify any anomalous network communication. \\ \hline
TCPXtract & It extracts files through interception from a network traffic based on their signatures.\\ \hline
SiLK & It supports an efficient capture, storage and analysis of network data flow, along with supporting network forensics. \\ \hline
TCPReplay & It classifies previously captured traffic, rewrites the layers headers and replays the network traffic. \\ \hline
P0f & A Passive OS fingerprinting tool that captures incoming traffic from hosts to networks, and detects the presence of firewall. \\ \hline
Nmap & A  tool that is used for OS fingerprinting and port scanning.\\ \hline
Bro & A network intrusion detection system that passively monitors network traffic. \\ \hline
Snort & A Network intrusion detection/prevention system that performs packet logging, sniffing and real-time traffic analysis. \\ \hline
\end{tabular}
\end{table*}

\subsection{Cloud Forensics}
The evolution of the digital world, led to the cooperation and collaboration between cloud forensics and digital forensics. In this context, cloud forensics~\cite{ruan2011cloud,cruz2012basics} play a key role in the big data era. Smart-phones, computers, laptops, tablets, and vehicles store their data in the cloud, which presents several benefits. These benefits include the scalability, the large capacity, and the on-demand accessibility. However, transferring the data over the network exposes it to various attacks against cloud-related domains. As a result, cloud users would become victims of cyber-crimes. Hence, this calls for digital forensics to be applied in the cloud environments (known as cloud forensics)~\cite{vlachopoulos2012model}. However, cloud forensics investigation is not a straightforward task, due to the difficulty in locating and identifying the evidence’s source, along with the lack of accountability of cloud providers. % in case of any given attack. 

\subsubsection{Cloud Forensics Tools}
 Despite the fact that there are few available cloud forensics tools, there is an increasing demand to develop more sophisticated and more efficient ones~\cite{mohite2015design}. In~\cite{krutz2010cloud,dykstra2013design,naazcomparative}, different cloud forensics tools are discussed and compared. Hence, the main cloud forensics tools are presented in \tablename~\ref{tab:5}.

\begin{table*}[!ht]
\centering
\caption{\textbf{Cloud Forensics Analysis Tools}}
\label{tab:5}
\begin{tabular}{|p{4 cm}|p{12cm}|}
\hline
\centering
\textbf{Cloud Forensics Tools (ClFTs)} & \textbf{Description} \\ \hline
FROST & It acquires data from API logs, virtual disks \& firewall logs to carry out digital forensics investigations, along with storing data logs in Hash trees whilst returning it in a Cryptographic form.\\ \hline
UFED  & A cloud Analyzer that allows forensics investigators to have a potential evidence for their investigations from cloud information sources, which speed up investigations, where cloud data/metadata is collected using UFED PRO to pack and use it for forensics examinations.\\ \hline
\end{tabular}
\end{table*}

\subsection{E-mail Forensics}
The number of email accounts and messages is growing constantly~\cite{wendt2016hierarchical}. In fact, e-mail forensics can be somehow related to computer forensics~\cite{maras2015computer}, given that the e-mails present the main platform of communications that ensures both confidentiality and integrity of the data shared. 

%\subsubsection{Email-Forensics Communication Platform}
%Moreover, Digsby creates multiple sessions, which challenges the forensics investigation process, since the forensics investigators cannot trace the log files stored at different locations in a given system. . In fact, using the DigLA forensics tool allows the analysis of system Random Access Memory (RAM), whilst finding the login credential of Digsby tools. Digsby is a communication platform made by researchers for emails, instant messaging and social networking. Hence,.

\subsubsection{Email-Forensics Approaches}
Many e-mail communication services (E.g outlook, yahoo, G-mail), platforms and marketing services (E.g Constant Contact, SendinBlue, AWeber, GetResponse and Mailchimp) are now being used. As a result, e-mail forensics are becoming key targets for hackers to retrieve companies confidential and sensitive information through phishing~\cite{jakobsson2005modeling} and spear phishing~\cite{parmar2012protecting}, whaling, war-dialing and vishing. This is due to targeting employees and staff, administrators and Chief Executive Officers (CEOs). Therefore, it is essential to find the right forensics tool to ensure the email forensics protection. In~\cite{paglierani2013towards}, Paglierani et al. managed to discover G-mail account credentials. This method consisted of the re-establishment of the already existing G-mail sessions, by employing an Email Forensics Extensible Markup Language/Email Forensics Resource Description Framework (EFXML/EFRDF) representation~\cite{paglierani2013framework} of email headers. In~\cite{kumari2016insight}, Kumari et al. used Digsby~\cite{yasin2013digla} for e-mail forensics investigation, where Digsby Log Analyser (DigLA) was issued to locate and detect data log files without the need to know details about the file location~\cite{yasin2013digla}. Furthermore, such a tool ensures a faster digital evidence collection for analysis purposes, whilst providing password decryption for machine specific and portably Digsby installations~\cite{yasin2013digla}.

\subsubsection{Email-Forensics Tools}
Different e-mail forensics tools were presented along with many other approaches and explanations in ~\cite{carrier2002open,stolfo2006behavior,devendran2015comparative}. Consequently,the  e-mail forensics tools are summarized in \tablename~\ref{tab:6} as follows.

\begin{table*}[!ht]
\scriptsize
\centering
\small
\caption{\textbf{ E-mail Forensics Investigation Tools}}
\label{tab:6}
\begin{tabular}{|p{3.5cm}|p{12.5cm}|}
\hline
\centering
\textbf{E-mail Forensics Tools (EFTs)} & \textbf{Description} \\ \hline
MailXaminer forensics email analysis Software
& It performs data collaboration \& access, and supports multiple mailboxes, equipped with search filters for an accurate forensics email search, recovery, along with storing \& preserving email evidence. \\ \hline
Aid4Mail Fookes software & An email archiving software that is used to search through mail, to produce results based on filters, and to convert mail to industry-standard production formats in a  highly accurate manner capable of safely preserving hidden metadata.\\ \hline
MxToolBox Email Software & It analyses IP addresses \& explores cyber-crime. \\ \hline
Paraben Email Examiner & It allows users to perform email examinations.  \\ \hline
OSForensic Software& It performs email forensics searches. \\ \hline
Free EDB Viewer & It is easy to use, with all user related mailbox information will be available here, and allows the offline viewing of EDB emails. \\ \hline
Exchange EDB Viewer & It can be associated with server exchange, and it contains crucial information to enable a forensics investigation.\\ \hline
FreeViewer EML Viewer& It uses EML format to store emails locally. \\ \hline
FreeViewer OST Viewer& It aids in viewing all email related information of an  OST file attached with Outlook. \\ \hline
FreeViewer PST Viewer& It reads the PST file content such as emails, attachments along with header information. \\ \hline
FreeViewer MSG Viewer& It reads the content within the Outlook MSG file along with headers. \& attachments information\\ \hline
SQL MDF Viewer & It views MDF Databases without SQL Server environments. \\ \hline
SQL LDF Viewer& It helps analysing what happened with SQL Server databases. \\ \hline
FreeViewer MBOX Viewer& It is used by many email clients, and can be easily viewed by the MBOX Viewer. \\ \hline
Opera MBS Viewer& It is a mail client that stores \& sends users emails, also generates MBS files. \\ \hline
DBX Viewer & The users can use this tool to view email, header information \& attachments. \\ \hline
WAB Viewer & It stores contact details in WAB format, and allows the retrieval of contact information stored in a WAB format. \\ \hline
ZDB Viewer & Zimbra can be connected to Outlook to view the email head information. \\ \hline
\end{tabular}
\end{table*}

\subsection{Malware Forensics}
Is also known as malware analysis that aims to study the process of determining the functionality of a given malware, along its impact and origins ad type (i.e virus, worm, trojan horse, rootkit, or backdoor). Since most mobile forensics analysis focus on the process of data acquisition~\cite{lessard2010android,hoog2011android}, it is very essential to identify suspicious applications, especially since malware can be hidden in malicious applications that seem to be legitimate, especially on Android platforms~\cite{zhou2012hey,di2010detection}.

%In fact, the message digest is an important cryptographic feature, where a normal application database is built by collecting the information message digest. However, permissions are required for Android programs due to the Android OS design. Thus, applications will be urged to apply for a list of permissions before being installed, which leads to suspicious permissions requirements that permit the detection of Android malware~\cite{zhou2012hey,di2010detection}.

%the malware analysis can be . a given . and.

\subsubsection{Malware Forensics Approaches} 
To ensure a better malware analysis, Various approaches were presented as part of solving malware forensics issues and challenges. In~\cite{li2012android}, Li et al. identified the first steps of mobile forensics analysis, which is based on identifying malicious applications. However, there still are many challenges related to the malware detection tools~\cite{schmidt2008enhancing}. In~\cite{khurana2010smart}, Khurana et al. divided the malware analysis into two main areas including behavioral analysis and code analysis. More precisely, the behavioral analysis aims to examine how a malware interacts with its environment~\cite{shukla2008application}, while the code analysis examines the malware malicious code~\cite{bayer2006dynamic}. Moreover, several methods and taxonomies were presented in~\cite{nicholson2012taxonomy}. However, the behavioral analysis still remains as an open issue. In~\cite{cook2016attribution}, Cook et al. relied on six main individual metrics to measure the level of the attribution’s effectiveness in the Industrial Control Systems (ICS) context applied to Critical Infrastructures (CI). In~\cite{rathnayaka2017efficient}, Rathnayaka et al. presented a malware analysis that integrates the malware static analysis with forensic analysis of memory dumps. Their approach can analyse advanced malware types that hide their behaviours without showing any possible artefact with an accuracy result of 90\%. However, this approach isn't compatible with different environments and suffers from memory dump size issues. 

%a malware analysis integrating a malware static analysis with forensics analysis of memory dumps. The advantage of this approach is the ability to analyse advanced malware types that hide their behaviours without showing any possible artefact. Simulation resulted revealed a 90\% classification rate which can also be maximized using different approaches in future experiments. However, this approach was not compatible with different environments. Another limitation is related to the memory dump size which may exceed the limit of 1 GB.

\subsubsection{Malware Forensics Tools}
Malware forensics tools are essential for any malware analysis and investigation. For this reason, different malware forensics tools are summarized in the following table \tablename~\ref{tab:7} and are further explained and discussed in~\cite{torres2015building,rastogi2013droidchameleon,cohen2011distributed,ligh2010malware}.
%However, the main malware forensics tools are summarized in the following table \tablename~\ref{tab:7}.

\begin{table*}[!ht]
\centering
\scriptsize
\caption{\textbf{Malware Forensics Analysis Tools}}
\label{tab:7}
\begin{tabular}{|p{3.5 cm}|p{12.5cm}|}
\hline
\centering
\textbf{Malware Forensics Tools (MFTs)} & \textbf{Description} \\ \hline
FOR610 & It explores malware analysis tools \&  helps forensics investigators, incident responders to acquire the needed practical skills to examine malicious programs that infect Windows systems. \\ \hline
Cuckoo~\cite{mosli2017behavior} Sandbox& An open source platform that automates malicious file analysis with a detailed meaningful feedback, and ensures malware detection \& protection. \\ \hline
Yet Another Recursive Acronym (YARA) Rules & An open source malware attribution tool that is used to classify \& analyse malware samples based on textual or binary patterns, malware are described based on their patterns. \\ \hline
Google Rapid Response (GRR) & An incident response framework that analyses specific workstations for malware footprints, consists of an agent deployed on the target system \& an interactive server infrastructure to interact with the agent, various forensics tasks can be performed on the client machine.  \\ \hline
Remnux & It uses the one-stop-shop approach to reverse engineer malware samples, help investigating browser-based malware, ensure memory forensics, and analyse multiple malware samples. \\ \hline
\end{tabular}
\end{table*}

\subsection{Memory Forensics}

Memory forensics reveal most of the crime credential information. This forensics type is meant to evaluate the physical memory including completeness, correctness, speed and amount of interference. As a result, various memory forensics useful tools and steps were listed  for an enhanced memory forensics investigation in~\cite{inoue2011visualization,vomel2011survey}. 

\subsubsection{Memory Visualisation}
%In fact. volatile memory forensics play a crucial role in a given digital investigation..
Volatile memories are characterized by their high speed of read/write operations. Thus, the visualisation is crucial in the volatile memory case, especially with large disk size and full disk encryption. In this context, MAC memory readers permit the visualization of the physical memory mapping, similarly to the show boot memory map used in the apple kernel debug kit~\cite{ford1997flux,inoue2011visualization}. 
\\
After mapping the memory, the investigation process requires a recording of the "correct and complete physical memory" of a given device or Basic Input/Output System (BIOS)~\cite{arbaugh1997secure,mihm2010system}. Hence, its main role is to record the image memory to avoid memory alteration, especially in the physical RAM~\cite{sikorski2012practical}. 

%\begin{itemize}
%\item \textbf{Physical Memory Evaluation:}
%After mapping the memory, the investigation process requires a recording of the "correct and complete physical memory". In fact, “complete” refers to the allocated and the unallocated physical address spaces of a given device and Basic Input/Output System (BIOS)~\cite{arbaugh1997secure,mihm2010system}. Moreover, “correct” means that a device’s page physical address space should be similar to the real address space of the main memory page. Thus, the physical memory evaluation tool has the role of recording the image memory to avoid memory alteration. In case of any data loss, forensics investigators proceed with different procedures for reconstructing it. The initial step is to seek and find the physical RAM, especially when attackers employ malicious software~\cite{sikorski2012practical}. 
%\end{itemize}

\subsubsection{Memory Forensics Approaches}
%. In~\cite{al2018live}, Ziad et al. reviewed the memory forensics approaches, along with the employed post-mortem analysis. This allows.  In fact, Random Access Memory (RAM) is a vital information source to take legal actions against cyber-criminals~\cite{harichandran2016cufa}. Therefore, it is designed to operate on unused memory space to maintain the RAM memory.This taxonomy can be deployed post or pre-incident to generalise memory access privilege model of recent hardware architectures.
As for memory forensics, various approaches were presented in~\cite{al2018live} to mitigate volatile memory issues (i.e Random Access Memory (RAM)~\cite{harichandran2016cufa}) to allow law enforcement agencies to take legal actions against cyber-criminals. In~\cite{shosha2013digital}, Shosha et al. developed a prototype to detect programs being maliciously used by criminals. This prototype is based on evidence deduction, which in turn, is based on traces of the program’s suspect. In~\cite{al2018live}, Olajide et al. used RAM dumps to extract information about the user input from Windows applications. In~\cite{chan2009framework}, Ellick et al. introduced a RAM forensics tool known as ForenScope~\cite{chan2010forenscope}. This tool ensures the investigation of a given machine by using regular bash-shell, which permits to disable the anti-forensics tools whilst searching for any potential evidence. In~\cite{stuttgen2015acquisition}, Johannes et al. presented a different approach. This approach investigates the firmware along its components, and improves forensics imaging based on page table mapping and PCI introspection. In~\cite{shashidhar2015digital}, Shashidhar et al. presented an approach that was aimed to target the potential value of a prefetch folder, along with the prefetch folder itself. This is used to start-up and speed-up a window machine program. In \cite{latzo2019universal}, Latzo et al. surveyed the memory forensics domain and presented a forensics memory acquisition taxonomy that is independent of the Operating System (OS) and the Hardware Architecture (HA), and can also be deployed pre/post-incident.

%\item \textbf{Locating Evidences:} 
%is highly important, especially after identifying the source. In fact, the user’s data might be stored in different locations, which may impose limitations and restrictions, due to the law differences between the different storage locations. 

\subsubsection{Memory Forensics Tools}
This paper summarizes the main different memory forensics tools that are presented and explained in~\cite{amari2009techniques,tabona2002top}. For example, PTFinder~\cite{schusterptfinder}, which is a forensics toolkit, was used to allow the investigation of the main memory content. For further clarification, these tools are summarized In the following \tablename~\ref{tab:8}. % summarizes these tools.  

\begin{table*}[!ht]
\scriptsize
\centering
\small
\caption{\textbf{Memory Forensics Investigation Tools}}
\label{tab:8}
\begin{tabular}{|p{3.5cm}|p{12cm}|}
\hline
\centering
\textbf{Memory Forensics Tools (MFTs)} & \textbf{Description} \\ \hline
SANS Investigative Forensics Toolkit (SIFT) & It includes all the needed tools to conduct an in-depth forensics or incident response investigation.  \\ \hline
CrowdStrike CrowdResponse & It is used as part of an incident response scenario to gather contextual information, and it can also scan host for malware detection.\\ \hline
Volatility & It is a memory forensics framework for incident response and malware analysis that allows the extracting of digital artefacts from volatile memory (RAM) dumps. \\ \hline
Sleuth Kit & It can be used to perform in-depth analysis of various file systems. \\ \hline
FTK Imager & It allows the examination of files and folders on local hard drives, network drives, CDs/DVDs, whilst reviewing the content of forensics images or memory dumps. \\ \hline
Linux ‘dd’ & It is used for a forensics wiping a drive whilst also creating a raw drive image.  \\ \hline
Computer Aided INvestigative Environment (CAINE) & It includes a user-friendly GUI, semi-automated report tools for Mobile \& Network forensics, along data recovery. \\ \hline
ExifTool & Fast \& It supports large range of file format,and it is used to read, write or edit file metadata information.\\ \hline
Free Hex Editor Neo & It is designed to handle \& load very large files, along, information gathering, or searching for hidden data. \\ \hline
Bulk Extractor & It scans a disk image, file, or directory of files and extracts information including e-mail addresses, URLs, and ZIP files.\\ \hline
DEFT  & It aims at helping with Incident Response, Cyber Intelligence and Computer forensics scenarios. \\ \hline
LastActivityView & It allows the view of what actions were taken by a user and what events occurred on the machine, useful to prove that a given user performed an action they denied. \\ \hline
DSi USB Write Blocker & It ensures a write-block to USB devices to keep data and metadata safe. \\ \hline
FireEye RedLine & It performs host memory \& file analysis, collects memory information \& gathers file systems to build an overall threat assessment profile.\\ \hline
PlainSight & It allows users to perform digital forensics tasks, along with examining physical memory dumps \& more.\\ \hline
HxD & It performs a low-level editing \& modification of a raw disk or main memory (RAM). \\ \hline
HELIX3 Free & It is built \& used in incident response, computer forensics and E-discovery scenarios. \\ \hline
USB Historian & It displays useful information including USB drives \& the serial numbers, to understand if data was stolen, moved or accessed. \\ \hline
\end{tabular}
\end{table*}

\subsection{Mobile Forensics}
Mobile forensics are classified as a new branch of digital forensics, which is about analyzing mobile devices to retrieve and recover digital data serving as evidences. This is done by preserving the integrity of the evidence in a un-contaminated and un-altered state~\cite{casey2010introduction,androulidakis2012mobile}. Mobile forensics tools and techniques rely on quantitative analysis approaches~\cite{marturana2011quantitative}. More precisely, this is due to the fact that mobile devices contain large amount of digital data and information (i.e contacts, call logs, SMSs, Wi-Fi information, IP/MAC address, Global Positioning Systems (GPS) signals, Bluetooth, etc...) that can serve as evidence.

%a variety of applications data, along with  the channels' information, internet connection details (service providers), Wi-Fi signals, Global Positioning Systems (GPS) signals, Bluetooth and radio signals. 
%Moreover, it includes taped conversations, digital pictures, audios and videos, e-mails and contact list. Therefore, it is essential to respect the standardized guidelines used to ensure a mobile forensics investigation by preserving the evidence's integrity and its user's privacy, so it can be safely stored, and copied to avoid it being damaged or deleted. Thus, resulting into a legal evidence creation, preservation and also presentation. 
\begin{itemize}
%in mobile forensics, locating evidences could be stored or hidden in.
\item \textbf{Locating Evidences:}
Locating evidences in mobile forensics is not an easy task, but it can be achieved.
In~\cite{chernyshev2017mobile}, Chernyshev et al. described the mobile phone evidences sources specifications. This includes the uniqueness and persistence of the devices identifiers, as well as the network information, and personal local settings (i.e saved passwords, cookies, electronic documents, web-browsing activities, etc...). 

%In fact, locating evidences also includes going through phonebook, contacts, calendars, alarms, emails, schedules, electronic documents and web-browsing activities, along with saved passwords and cookies. In fact, more evidences can be found in saved messages, recorded calls, archives, along with audio messages, images and video calls history. Moreover, the GPS function can reveal the accurate location of the device, as well as the installed applications including social media application (Facebook, Twitter, etc..) and instant messaging application (WhatsApp, Viber, etc.). Furthermore, cloud services are included through a device scheduled backup. This helps accessing and identifying local and personal networks through the reliance on Wi-Fi, 3G, 4G, 5G, Long Term Evolution (LTE) along with their extensions, as well as Bluetooth, and the mobile telecommunications carrier cell tower. Each one of them can serve as a vital evidence or/and clue that helps forensics investigators to proceed with their forensics investigation. Moreover, the running firmware, which is the OS, and the hardware components are used also to extract and locate the evidences.

\item \textbf{Digital Artifacts:}
Before describing and classifying the mobile forensics tools, it is important to know what digital artifacts can be retrieved first, including their types. For this reason, \figurename~\ref{fig:md} was presented.
%As a result, these artifacts are divided and classified into three main types (see \figurename~\ref{fig:md}): 
\begin{figure*}[!ht]
  \centering
    \includegraphics[scale=0.5]{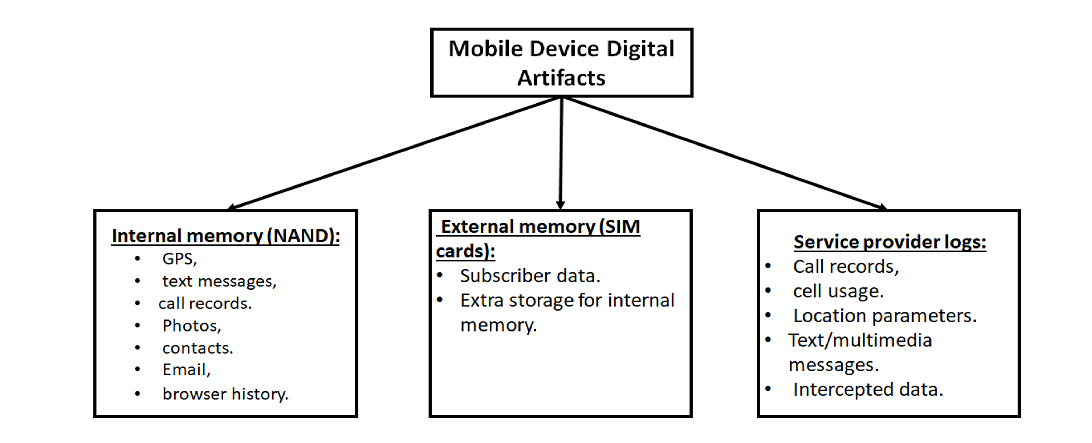}
    \caption{Mobile Forensics Digital Artifacts}
    \label{fig:md}
\end{figure*} 

\begin{itemize}
\item \textbf{Internal Memory:} includes the NAND flash memory \cite{lee2002effects}. In such a memory type, many evidences can be found, such as digital data, SMS, call logs and browser history.
%GPS location, text messages, call records, call logs, contact list/number, e-mails, applications, digital data (audio, video, text, images), SMS and browser history.
\item \textbf{External Memory:} includes the SIM card, where several evidences can also be found and retrieved, such as the subscriber's data, real-time location, and additional internal memory storage. 
\item \textbf{Service Provider Logs:} includes call logs, duration and usage that can be identified and retrieved even after deletion. 
%Moreover, location parameters, text and multimedia messages along with the intercepted data can be retrieved to serve as a legal evidence at courts.
\end{itemize}

\item \textbf{Mobile Forensics Tools}
%researchers consider mobile forensics to . . For this reason, they were considered as.
In recent years, mobile devices were involved in both crimes and cyber-crimes scenes alike as vital key digital witnesses to investigate the crimes involving mobile devices. For further technical explanation, \tablename~\ref{tab:9} summarizes the main mobile forensics tools.

\end{itemize}
 
\begin{table*}[!ht]
\centering
\small
\caption{\textbf{Description of Mobile Forensics Tools}}
\label{tab:9}
\begin{tabular}{|p{3cm}|p{14cm}|}
\hline
\centering
\textbf{Mobile Forensics Tools (MFT)} & \textbf{Description} \\ \hline
OXYGEN forensics KIT & It extracts data from mobile devices and analyses it, and ensures an efficient use and acquire of information \\ \hline
Pilot-Link & It is an open source software designed for the Linux client, and provides the means to find the logical device’s contents which can be examined manually by palm OS emulators~\cite{jansen2005overview} \\ \hline
Encase & It allows palm OS device, and stores the physical bit stream image of file for future use \& can be viewed any time in the future, also it helps snapping physical and logical snapshot of the device’s current state. ~\cite{grover2013android}\\ \hline
PDA Seizure & It ensures a logical access to information through the use of API protocol to allow desktop applications to communicate with mobile devices \& allows investigating the pocket PC and palm OS, along bookmarking, graphical libraries.  \\ \hline
XRY & It is used to analyze and recover crucial information from mobile devices, and it is made up of hardware device and software \& designed to recover data for analysis. \\ \hline
PlainSight & It recovers deleted data (calls, SMS, images etc..) from all kind of smartphones including Android, iPhone and BlackBerry. \\ \hline
Cellebrite UFED & It is a unified workflow that allows investigators along first responders to collect, protect and act on mobile data in a fast yet accurate way without risking data being compromised. \\ \hline
\end{tabular}
\end{table*}

\begin{figure*}[!ht]
  \centering
    \includegraphics[scale=0.4]{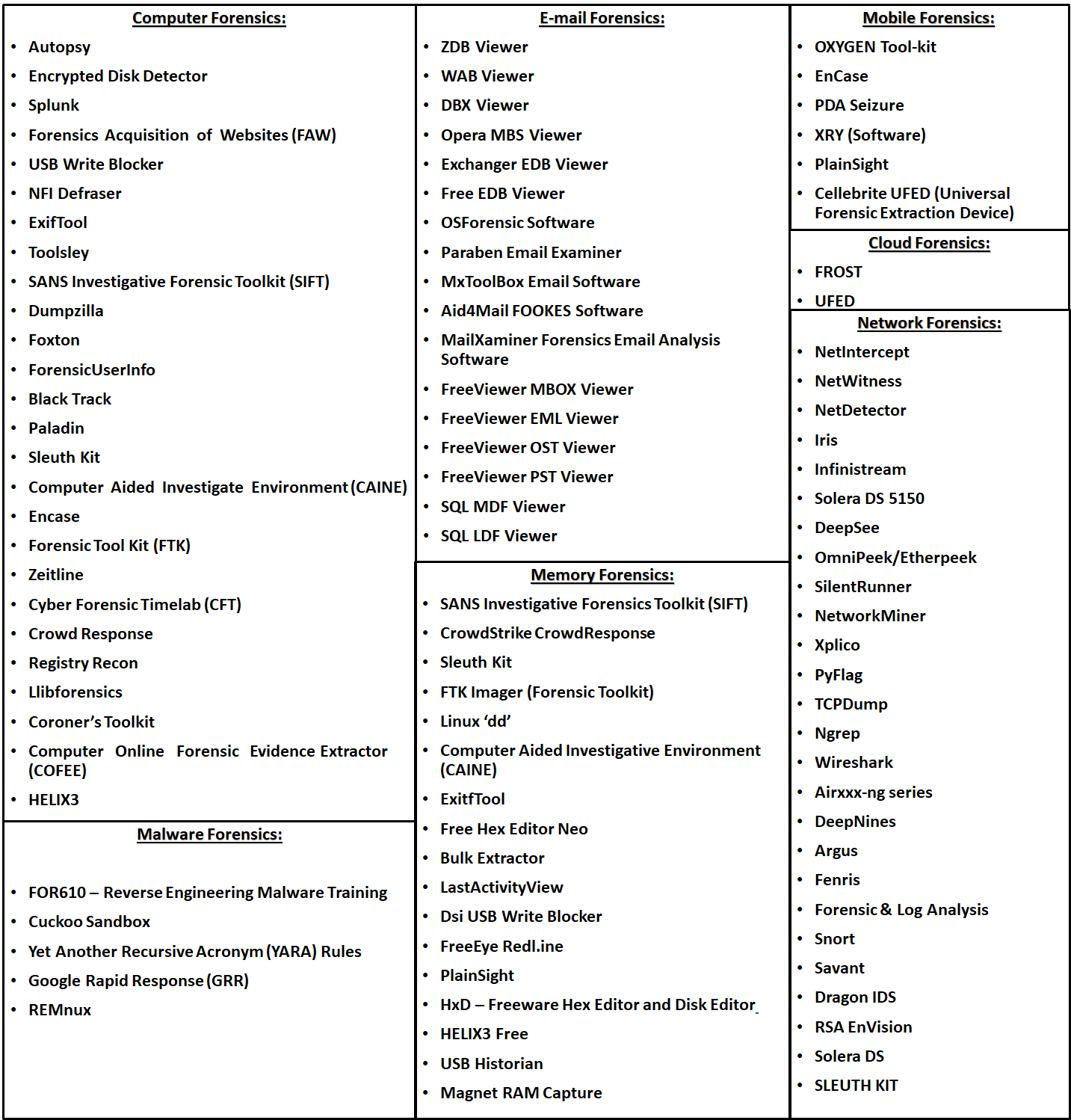}
    \caption{Classification of Forensics Tools}
    \label{fig:q2}
\end{figure*}

As a result, these forensics tools are summarized in  \figurename~\ref{fig:q2}.

%    \item \textbf{ :}   The sumo wrestlers refused to turn over their mobile devices to law enforcement claiming their phones were damaged due to water or the battery had died in the phones.

\subsection{IoT Forensics} 
IoT is a smart network capable of decision-making and self-managing, whilst being linked to various domains including medical IoT \cite{yaacoub2019securing}. Unlike traditional forensics that rely on the Triple-A domain (known as Authentication, Authorization and Accounting)~\cite{glass2000mobile}, IoT forensics rely on Radio Frequency Identifiers (RFID) tags, as well as the network and sensor nodes. In fact, \tablename~\ref{tab:4} contains a comparison between traditional forensics and IoT forensics.
%Its impact on the forensics world seems to be alarming. This is due to its large scale employment, evidence sources and types, its supported protocols,its big generated data, and its supported network protocols. 

\begin{table*}[!ht]
\centering
\small
\caption{  {Traditional Vs IoT Forensics}}
\label{tab:4}
\begin{tabular}{|l|l|l|}
\hline
\centering
 \textbf{Comparison} & \textbf{Traditional Forensics}&  \textbf{IoT Forensics} \\ \hline
 {Ownership}&  {Individuals, companies, governments, etc..} &  {Individuals, companies, governments} \\ \hline
 {Protocols}&  {Ethernet, Wireless, Bluetooth, IPv4, IPv6} & {Wireless, Bluetooth, Zigbee, RFIDs}\\ \hline
 {Data Size}&  {Terabytes} &  {Exabytes} \\ \hline
 {Number of Devices}&  {Billions} &  {40+ Billions}\\ \hline
 {Networks Nature} &  {Wired, Wireless, Bluetooth, GSM} & {Wireless,  Bluetooth, RFID, WSN, 4G/5G} \\ \hline
 {Evidence Source}&  {AAA, gateways, social networks} &  {RFID tag/reader, sensor nodes/networks} \\ \hline
 {Evidence Type}&  {Electronic documents, standard file formats} &  {Any available format} \\ \hline
\end{tabular}
\end{table*}

In the following, the IoT characteristics are detailed based on the comparison elements included in \tablename~\ref{tab:4}.
\begin{itemize}
\item \textbf{Evidence Source:}
Identifying the source of evidence requires having the knowledge of the type of devices being in use (i.e Software, Hardware and OS). Moreover, it also requires collecting the necessary forensics evidences from IoT-based and digital cyber-crime scenes. %This includes the data and logs found in victims' personal computers, laptops, tablets, smart-phones and other IoT devices, including usernames, e-mails, passwords, call logs, files, images, videos, audios, texts and so on.
%\item \textbf{Number of Devices:}
\item \textbf{Crucial Information:} 
due to the increasing growth of the devices numbers, with more than trillions of interconnected devices being operational on IoT networks~\cite{dunkels2007rime}, the aim is to locate and identify any available information that proves to be crucial for a given forensics investigation, despite the explosions in terms of data size on IoT platforms~\cite{coetzee2012inclusion} which may reach up to 40,000 Exabytes by 2020~\cite{gantz2012digital}.
%is being available and stored on these devices. This data is of high value
%The number of devices is on an extreme constant growth. In~\cite{dunkels2007rime}, Dunkels et al. stated that there are trillions of interconnected devices in the IoT network, which outnumbers those in the traditional case. What really matters is the crucial information that is being available and stored on these devices. This data is of high value to ensure their digital forensics investigations.
%\item \textbf{Data Size:} 
%In IoT, there is a data explosion due to the increased number of interconnected devices. These devices are constantly and continuously communicating and exchanging real-time data and information across the various IoT platforms and services~\cite{coetzee2012inclusion}. In fact, it is about to reach up to 40,000 Exabytes by 2020~\cite{gantz2012digital}.
\end{itemize}

\subsection{IoT Persistent Issues} 
%The developed mobile forensics tools become obsolete due to the emergence of new devices. This is also caused by the evolution of mobiles' OS with new encryption techniques. 
One of the main IoT forensics issues is the lack of a reliable IoT forensics application~\cite{watson2016digital}. Moreover, there is no existing digital forensics guidance that allows retrieving data from an IoT device, in case of an active forensics investigation, or an occurring cyber-event. More precisely, the embedded technologies are challenging due to their reliance on traditional computer OS or even magnetic data. Therefore in~\cite{watson2016digital}, Watson et al. introduced the need for an advanced data recovery technique whenever a data acquisition from an embedded remote IoT device is required. In fact, it seems like the digital forensics complexity is due to three main issues:
\begin{itemize}
    \item \textbf{Inaccessible Data Storage:} On-board data storage cannot be accessible through the use of traditional digital forensics methods.
\item \textbf{Dispersed Cumulative Datasets:} might exist in various yet different geographical locations. 
\item \textbf{Unreadable Data:} in case the data was acquired, the issue is that it cannot be readable or accessed with the available tools.
\end{itemize}

\section{Digital Forensics Challenges}
~\label{sec:3}
So far, there are plenty of challenges and issues that surround the Forensics domain as whole \cite{bennett2012challenges,luoma2006computer}. As a result, digital forensics challenges can be divided into technical, operational, legal and investigative challenges. This taxonomy is slightly similar to the on.e presented by Karie et al. in~\cite{karie2015taxonomy}.

\subsection{Technical Challenges}
During a forensics investigation, different types of technical challenges that require dealing with cryptographic and non-cryptographic data are encountered. This includes data size, data location, data hiding, data deletion, anti-forensics tools, and incompatibility, which may result into hindering an investigation, or extreme consumption of resources and time.
\begin{itemize}
\item \textbf{Cryptographic Challenges:}
The level of encryption plays a key role in a forensics investigation, since it can vary between symmetric and asymmetric encryption techniques. In fact, hackers and cyber-criminals use it to preserve the privacy of their data to avoid its capture. Hence, this explains their reliance on anonymity, homomorphic encryption, secret sharing and differential privacy, along other encryption mechanisms to make it almost impossible for digital forensics investigators to decrypt them.
\item \textbf{Data Size:}
Another technical challenge is related to the size of data (small data, medium data, big data~\cite{cardenas2013big,shalaginov2017cyber}) that requires to be retrieved. This also includes seeking which data can be used as evidence, by identifying what data is relevant and what data serves no purposes. Hence, cyber-criminals rely on covering their tracks and leaving data that serves no purpose to waste the investigators time.
\item \textbf{Data Location:}
Locating where the data is stored is yet another challenge, since, it is not easy to know where the data is stored and located. It is due to the fact that hackers use VPNs, proxies and TOR to perform their attacks anonymously without leaving any trail or trace back, which limits the amount of data being recovered and analysed for any possible trace or evidence.
\item \textbf{Data Wiping:}
Wiping or deleting data became also a serious challenge for forensics investigators, since hackers and cyber-criminals kept on deleting their data beyond recovery. Thus, leaving forensics investigators with little or no evidences at all to carry out with their digital investigation.
\item \textbf{Data Hiding:}
Hiding data is a popular technique used by both cyber-criminals and hackers alike. Such a technique relies on steganography to hide data. In some cases, hackers also rely on hiding their data in volatile RAMs (Random Access Memory). Therefore, once the power is off, the data is completely erased, and there is nothing that digital investigators can do to retrieve it.
\item \textbf{Anti-Forensics Tools:}
Anti-forensics tools are in use due to their popularity and effectiveness to counter forensics investigations and eliminate any source of evidence that can be retrieved or/and traced back. In fact, these tools impose a serious risk and threat to any digital investigation, since it is very easy to use them to erase data beyond recovery.
\item \textbf{Incompatibility:}
Due to the various techniques and technologies used by IoT devices, forensics tools are alsmost unreliable when it came to deal with the different types of devices, especially counterfeit devices. This makes any data's retrieval process very difficult and almost impossible.
\item \textbf{RAID:}
The use of Redundant Array of Independent Disks as a technology that combines different physical drives into a single logical unit resulting into a data storage virtualization~\cite{gul2017survey} is increasing. This technique mainly relies on arbitrary disk order, stripe order, stripe size and block size, along with the use of uncommon RAID controllers to eliminate any evidence, which is proving to be very difficult to recover them.
\item \textbf{Cloud Computing Storage:}
Due to the emergence of cloud computing~\cite{dahbur2011anti}, data is moved and outsourced to third parties. Hence, a new challenge will be thrown against a forensics investigation process especially with untrusted and semi-trusted third parties. Once the data is stored or transferred through cloud services, it can be transferred across different countries that impose different regulations. This would seriously complicate and affect a given investigation.
\end{itemize}

\subsection{Operational Challenges}
Aside technical challenges, operational challenges also present a serious threat to the forensics investigation process. This is due to the lack of incident management, lack of standardized procedures, and lack of forensics readiness.
\begin{itemize}
\item \textbf{Lack of Incident Management:}
Lack of incident management, is also known as lack of incident detection, response and prevention. In other terms, digital forensics investigators are still incapable of detecting any incident. In fact, even if they managed to detect an incident, they are either unable to respond to it in time, or they lack the ability to respond at all. Furthermore, there is lack of tools to prevent an incident from occurring, even with the reliance on IDS/IPS hybrid responses~\cite{aydin2009hybrid,garg2016hybrid,gupta2015hybrid}.
\item \textbf{Lack Of Standardized Procedures:}
Due to the lack of standardization of both procedures and policies, digital forensics investigators are facing real challenges in acting and reacting in the right way when an incident occurs.
\item \textbf{Lack of Forensics Readiness:}
Due to the lack of incident management and standardized procedures, forensics investigators severely lack of any sort of readiness to deal with a cyber-crime scene and retrieve forensics evidences. This makes it more difficult to detect and trace back any digital evidence.
\end{itemize}

\subsection{Legal Challenges}
After encountering technical and operational challenges, another type of challenges requires further attention to deal with and overcome. This issue is based on legal challenges which includes lack of jurisdiction, lack of legal process, security issues, insufficient support and privacy concerns ~\cite{khan2014forensic}.
\begin{itemize}
 \item \textbf{Lack Of Jurisdiction:}
Lack of jurisdiction is due to the lack of the official power to make legal decisions and judgements. This is caused by the tight human right constraints, which presents a serious challenge against forensics investigators to track down and arrest hackers depending on the type of their committed crimes.
\item \textbf{Lack of Legal Process:}
Lack of legal process includes the lack of any criminal prosecution by the court to take the necessary legal decision and judgment over a suspect that was proven to be guilty. Thus, lacking the knowledge of digital matters, with no firm laws being applied by courts to prosecute cyber-criminals.
\item \textbf{Security Issues:}
Security issues are part of the legal challenges, especially with the victims’ concerns towards trust issues. This also includes the accuracy and timeliness of the forensics investigation . More precisely, it is based on the level of trust that victims have in the federal services, whilst also providing them with details to track down and arrest cyber-criminals.
\item \textbf{Insufficient Support:}
Insufficient support is another challenge, consisting of the lack of funds, and the lack of public support. In fact, the lack of trust and support from the public can result into the lack of confidence in the job performed by the forensics investigators.

\item \textbf{Preserving Users \& Victims' Privacy:} the sharp rise of social engineering social-media based attacks is due to users’ excessive online life sharing aspects on social media. However, due to users’ privacy fears and concerns from forensics investigators breaching their privacy imposes a challenge, since an event and attack cannot be easily reconstructed without violating users’ privacy \cite{caviglione2017future}.
\item \textbf{Legitimation:} still remains a challenge due to the shifting from modern infrastructure to fog computing and third parties such as platform-as-a-service frameworks. Therefore, this offers a new complex and virtual issues. As a result, modern digital forensics investigations must be executed legally and without violating laws in the borderless virtual cyber-world \cite{caviglione2017future}.
\item \textbf{Responsibility:} due to social media platforms (Twitter, Facebook, etc..) turning a blind eye on fake news, this ensured their continuous spread  which led to various negative effects including violence, hatred, terror and fear. Moreover, this also led to the rise of phishing and privacy attacks by stealing users’ credentials for the purpose of blackmailing, forgery, fake identity or privacy breaches. Therefore, social media companies must allow and help forensics investigators to track down the source of fake news and prevent its spread by containing it and providing the right helpful information to locate the perpetrators and enhance \cite{caviglione2017future}.
\end{itemize}

\subsection{Investigative Challenges}
Investigative challenges are usually caused by the lack of qualified forensics personnel, and the lack of forensics knowledge when it comes to use forensics tools. 
\begin{itemize}
\item \textbf{Interoperability of Forensics Tools:} since forensics tools store data in various different format types which vary between different databases, datasets and data structures types, this still remains a real challenge and issue~\cite{barmpatsalou2018current}. The lack of standardization and uniformity, makes these digital forensics heterogeneous by nature. Therefore, there’s urgent and persistent need for a unified data format for the acquired forensics digital data.
 \item \textbf{Lack of Qualified Forensics Personnel:}
Lack of qualified digital forensics personnel is a challenge in itself, especially with the lack of training and experience in the forensics field. In fact, this is due to the lack of education, where many digital forensics investigators operate without obtaining any official forensics certificate.
\item \textbf{Lack of Standardized Threshold:}
The lack of standardized threshold is due to the lack of issued certificates for forensics investigators that allows them to be classified as authorised. Many investigators claim to be forensics investigators due to the fact that they literally know  details, or have an experience in the forensics domain. Therefore, the lack of a standardized threshold to classify forensics investigators remains a consistent challenge.
\item \textbf{Lack of Forensics Knowledge:}
Despite the lack of experience, knowledge and skills, another challenge is the lack of forensics tools and kits. Moreover, in most cases, forensics investigators are incapable of using these tools or these forensics kits due to the lack of expertise and skills. This might result into the loss, or damage of original data beyond recovery.
\item \textbf{Lack of Forensics Investigative Skills:}
Another challenge is the lack of investigative skills. In fact, these skills can be classified into soft investigative and hard investigative skills. 
\end{itemize}

\subsection{Mobile \& Device Challenges}
Smart and mobile devices along with computers, laptops, and tablets are part of the IoT world. Thus, forensics investigators encounter many challenges when extracting data from these devices ~\cite{sai2015forensic, jadhav2016forensic}. These challenges are listed in the following:
\begin{itemize}
\item \textbf{Heterogeneous Nature:}
The heterogeneous nature of mobile, digital and IoT devices, especially with different hardware and software configurations and components~\cite{gronli2014mobile}, is challenging for digital forensics investigation. Therefore, different forensics techniques and tools are needed to investigate and disassemble a given device to prevent the risk of destroying data.
\item \textbf{Built-in Security Features:}
Built-in security features are capable of  limiting the access to any device. These features are related to authentication, identification and verification. Moreover, the use of biometrics imposes a serious challenge to the forensics investigation process.
\item \textbf{Lack of Forensics Tools:}
In fact, there is a lack of forensics tools, kits and equipment that can be used for IoT devices forensics investigation. Therefore, there is a limited chance of ensuring that data can be retrieved safely and carefully without risking damage or destruction. In fact, the existing forensics tools are incompatible with the emerging IoT devices. 
\item \textbf{Malicious Applications:}
Malicious applications are used by cyber-criminals to perform surveillance attacks. Once these applications are installed, a Trojan or worm will be activated on a given device, capable of ensuring spyware, ransomware, botnet or even DoS attack. This offers the ability to delete, alter, modify and even manipulate the device’s data, whilst gaining an unauthorized privileged access. 
\item \textbf{CTI Challenges:}
Due to the existing exploitable vulnerabilities and security gaps in any given system, cyber-criminals may carry out their cyber-attack through infection and exploitation. In fact, attackers start using innovative methods to attack and target their victims, by relying on spear-phishing and social/reverse engineering techniques~\cite{conti2018cyber} . Moreover, such attacks can masquerade a given malware into a PDF file, image or even a video, that would run on the victim’s machine~\cite{elingiusti2018malware} without his knowledge. Therefore, this would lead to another form of backdoor~\cite{felt2011survey,young2003backdoor} to the victim's system. %Thus, proving itself to be a seriously challenge issue.
\item \textbf{Legal Limitations:}
Since mobile devices are part of the IoT world, in case of an international crime, different laws and different security measures can cause conflicts among different countries. In fact, it is due to the absence of a uniform jurisdiction and legal processing systems that can be applied to ensure a better cooperation and collaboration between different peers. 
\item \textbf{Devices Components:}
Device components can be divided into:

\begin{itemize}
\item \textbf{Software components:} are related to the use of different OSs and software (e.g Apple, Android, etc. on smartphones, Windows, Linux etc. on computers). Each OS operates differently, which presents a challenge for forensics investigators, since this requires different investigation approaches.
\item \textbf{Hardware components:} including the storage of data on volatile memories including RAMs, or on magnetic storage. Thus, the smaller the physical size of a given storage area, the harder for forensics investigators to investigate it without risking damaging it.
\end{itemize}
\item \textbf{Wireless Communications:}
It is also important to note that wireless communications impose a serious challenge to the forensics investigators, especially when dealing with well-trained and well-experienced hackers (cyber-(industrial)-espionage, advanced persistent threats) that cover their tracks aside the use proxies, VPNs and TOR~\cite{yen2012host,hoang2014anonymous,ramadhani2018anonymity} to hide their committed moves. This eliminates and reduces  possible evidences to avoid being detected and tracked down by forensics investigators. 
\item \textbf{Devices Types:}
Another challenge arose, especially with the huge number of counterfeit devices including laptops, PCs, smartphones and tablets being spread across the market. This is due to their cheap prices and their lack of security measures. This made it extremely easier for an attacker to use them to bait his victims to install fake applications, or use them to hide data and information. In many cases, these fake devices can serve as bots (zombies) and lead an anonymous attack on the attacker's behalf. This also includes their usage to easily logically destroy the data through the use of anti-forensics tools, or through physical destruction. Unlike original devices, it is harder for forensics investigators to track a genuine device compared to a counterfeit device.
\end{itemize}

\subsection{Big Data Challenges} 
Although the big data challenge was briefly mentioned before, it is important to explain it in more details to highlight its importance (see \figurename~\ref{fig:forensics Challenges}). Dealing with big data issues and challenges ~\cite{chang2015nist,chen2014big,madden2012databases}, requires extra efforts to achieve the intended results and hunt down cyber-criminals. In~\cite{adedayo2016big}, Adebayo classified these challenges depending on the data’s variability, velocity and volume. In fact, other challenges related to the accuracy and heterogeneity, validity, and trustworthiness are discussed in the following:

\begin{figure*}[!ht]
  \centering
    \includegraphics[width=\textwidth]{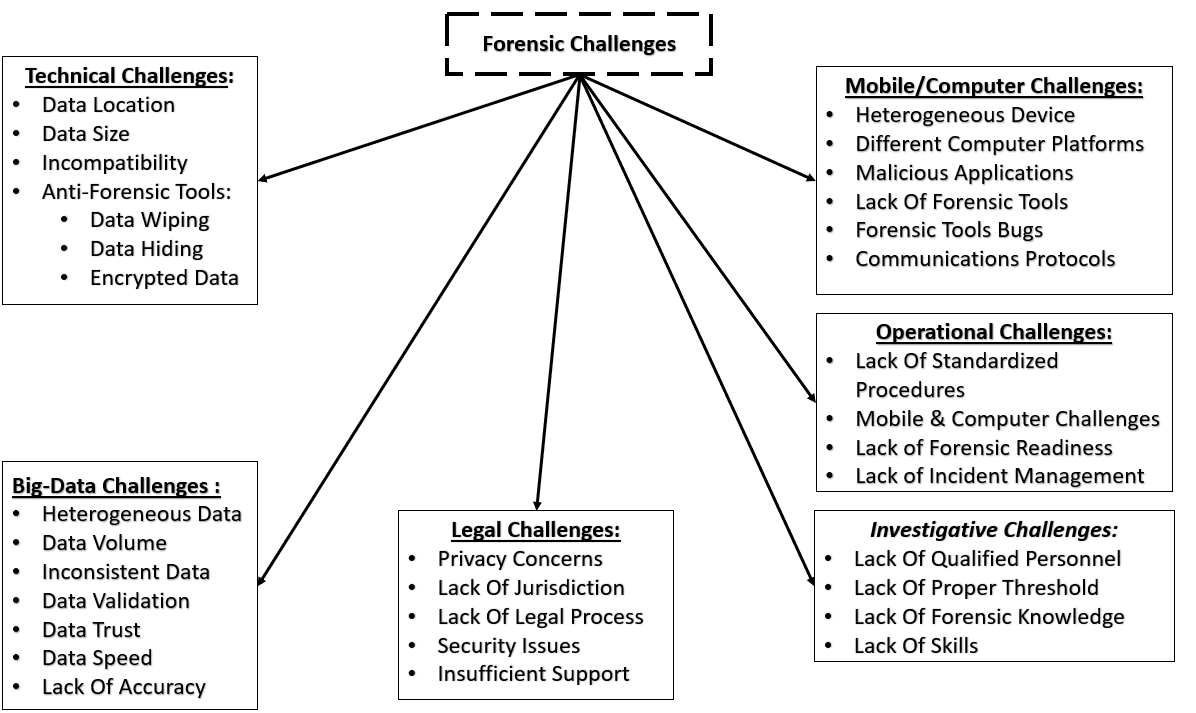}
    \caption{Forensics Challenges}
    \label{fig:forensics Challenges}
\end{figure*} 

\begin{itemize}
\item \textbf{Lack Of Accuracy:} 
Due to the big data size, an accuracy issue related to the nature, source and value of the retrieved evidence(s) arose. In most cases, big data offers zero or poor evidence. This is the reason why it presents a real problem for forensics investigators by wasting their time searching for any useful information. 
\item \textbf{Heterogeneous Data:}
data collected from different sources can either be structured, semi-structured, or non-structured. This presents a serious problem that can tackle down the forensics investigation. In case the data was non-structured, there's no format to support it properly, resulting into a waste of time and resources.
\item \textbf{Data Inconsistency:}
is related to the volume, velocity and variety of big data~\cite{adedayo2016big}. This presents an extra burden since cyber-criminals rely on a high volume of big data which is in most case irrelevant and inconsistent. Moreover, the nature of the retrieved data depends on whether it is structured or not-structured. This requires an additional time and resources to reconstruct the data and analyse it.
\item \textbf{Data Validation:}
Another challenge is related to the data validity, especially if dealing with metadata that serves for a short amount of time. In this case, the challenge is to see how long the data can survive, especially on volatile memories such as RAM.
\item \textbf{Data Trust:}
 Data hiding, manipulation, and alteration, make it difficult for forensics investigators to prove that the retrieved evidence is legitimate. This is due to the possibility that the collected data is modified or even altered. Therefore, it is challenging to prove the legitimacy of the retrieved evidence to be justifiable, legal and usable in courts. 
\item \textbf{Data Speed:}
Data velocity or data speed is related to the speed of which the data is processed at~\cite{adedayo2016big}. In other terms, it is the speed of generating or moving data around. In fact, big data velocity requires the need for data acquisition and analysis at a higher scale to maximize the data value. 
\item \textbf{Data Volume:}
can be defined as the amount of data generated, especially when dealing with big data, where a huge amount of data is generated. Consequently, this requires ensuring the scalability of data storage, along with the need for a distributed data processing approach. This presents a serious challenge for digital forensics investigators, especially if the data was hidden in networks, clouds and memories, or even encrypted.
\end{itemize}

\subsection{Educational Challenges:}	
Education also plays a key-role as the initial source of most of the occurring challenge, due to lack of training, experience, funding and available personnel. Potential forensics investigators must undergo excessive forensics studies, courses and training, while also being familiar with various forensics tools and aware of the most frequent anti-forensics activities. As a result, educational challenges are presented as follows:
\begin{itemize}
\item \textbf{Funding:} the lack of or little funding by government, organisations and enterprises renders the forensics domain very limited in terms of capability-wise and performance-wise. Hence, it still remains as a challenge \cite{jin2018game}. Therefore, more funding is needed towards researching and developing digital forensics fields to improve the collaboration among various forensics investigators and colleagues.
\item \textbf{Mindsets:} the mindsets of many individuals including numerous, universities, organisations and governments still believe that the digital forensics domain is still not effective. It is clear that there’s a huge funding, knowledge, and experience gaps in this field, with many attacks remaining untraceable and anonymous. Therefore, most military, law enforcement, police and governments mindsets are now shifting towards enhancing their forensics skills especially in the rise of cyber-terrorism \cite{bogdanoski2013cyber}, cyber-warfare \cite{hughes2017hierarchy,kosenkov2016cyber}, cyber-espionage \cite{langner2011stuxnet} and cyber-politics \cite{soriano2013internet} era. %Therefore, resulting.
\item \textbf{Liaison \& Communication Support:} the lack of support among communities is caused (aside funding) by the lack of proper discussion, communication and liaison. This results into conflict caused by a total lack of information or misinformation being shared. As a result, more collaboration and encouragement between communities (mainly universities and institutions) is required to share and enhance their forensics data sets for a better investigative outcome.
\item \textbf{Institutions Support:} the lack of institutions’ support including universities is primary related to the high cost of available education, tools, license and lack of skills to use them. Therefore, more fund, focus and education must be invested in this forensics field which may include national/international competitions, collaborations, opportunities and exchange of students \cite{jin2018game}, such as the "GenCyber" program \cite{luciano2018digital}.
\item \textbf{Standards Development:} the lack of communication and collaboration between different national and international forensics universities, facilities and organisations led to the creation of various software and hardware forensics tools that perform the same task of artifacts collection, categorisation and analysis. Therefore, proving to be a loss of time and resources alike. Hence, research communities need to work and agree on a unified set of standards and formats and abstractions~\cite{caviglione2017future} to avoid redundancy and collusion issues, with a focus on the timeliness of these standards, along their accuracy and effectiveness. 
\end{itemize}

\section{Anti-Forensics}
~\label{sec:4}

Cyber-criminals are now excessively using new sophisticated methods to perform their attacks. These methods are based on covering their tracks to avoid detection. This is achieved by using anti-forensics techniques and tools to alter and delete log and audit files~\cite{conti2018cyber}. As a result, Common Vulnerability Scoring System (CVSS)~\cite{petraityte2018model}, along with Static Malware Traffic Analysis (SMTA)~\cite{shalaginov2018machine} are not enough to mitigate this issue. Thus, anti-forensics presents a seriously threatening challenge for the IoT domain that heavily relies on cloud computing services to store and process big-data. Additionally, their use would drastically hinder the progress of forensics investigators by preventing them from carrying out their investigations. Hence, it is essential to overcome the existing challenges and limitations that Cyber Threat Intelligence (CTI) domains suffer from~\cite{pandya2018forensics}. Anti forensics are also known as counter-forensics~\cite{hausknecht2017anti}. Their task is to disrupt a given forensics investigation. Hence, different anti-forensics techniques, tools and approaches are being employed to evade detection and avoid being caught. This section presents and discusses them in details to help identify them and protect digital evidence(s) through mitigation and implementation of the right security measures. 

\subsection{Anti-Forensics Aspects}
Anti-forensics are used to remove, alter, disrupt or illegally interfere with the evidences found on digital devices in a digital/physical crime scene. Different anti-forensics aspects were discussed in~\cite{kessler2007anti}. This included the reliance on the Metasploit anti-forensics project~\cite{metasploit2007metasploit}, which is an open source project used to provide pen testing, Intrusion Detection Systems (IDS), information system exploit, and other services. Moreover, the Metasploit Anti-forensics Investigation Arsenal (MAFIA) has been used to improve the digital forensics processes, while also validating the forensics tools. MAFIA included the following components:
\begin{itemize}
\item \textbf{Transmogrify:} aims to overcome the EnCase’s file signature detection. It is done by masquerading a file into another file type. 
\item \textbf{Timestamp:}
as a program, it is capable of altering New Technology File System (NTFS) timestamp values. This is done through the MAC file entry modification, and entry update. These tools help confuse forensics investigators and further complicate their forensics investigation~\cite{hilley2007anti}.
\item \textbf{Sam Juicer:}
is a program that compromises the hashes of a given security access manager file. In fact, Sam Juicer runs over a memory/Image 
Local Security Authority Subsystem Service (LSASS) channel to store password hashes on a Windows system without leaving any trace or signature on the disk. Thus, avoiding the risk of being detected.
\item \textbf{Slacker:}
is a program that allows cyber-criminals to hide their data within a slack space found in the memory. This slack space is created when the file system (e.g. NTFS) allocates more space for a file, where the unused space is called a slack space. Therefore, this space forms a perfect place for data hiding~\cite{hilley2007anti}.
\end{itemize}

\subsection{Anti-Forensics Techniques}
As a result of the constantly increasing use of anti-forensics techniques, different anti-forensics approaches were presented to show how easy it is to target and tackle down a given investigation.
  In~\cite{peron2005digital}, Peron et al. discussed the attacker aims when using anti-forensics techniques and focused on how the attacker is capable of hiding, destroying, manipulating and/or even preventing the creation of any given evidence. In~\cite{wundram2013anti,kessler2007anti} Wundram et al. and Kessler et al. presented four main categories of anti-forensics approaches to provide artefact wiping, data hiding, trail obfuscation and other attacks against the computer forensics tools and the forensics investigation process. In~\cite{harris2006arriving}, Harris et al. presented a new method of data transformation which either hides, destroys, eliminates or counterfeits the evidence and its source. 
  In~\cite{garfinkel2007anti}, Garfinkel combines the attack targets and goals to present the already existing tools. In~\cite{wundram2013anti}, Wundram et al. presented the integration and harmonization of existing classification schemes into a single taxonomy with two dimensions respectively. The first dimension refers to the goal of the attacker who aims to avoid or delay the investigation. The second dimension refers to the target of the attacker which can be one of the following: the evidence, the forensics tool, or the investigator. In~\cite{stamm2012temporal}, Stamm et al. presented a temporal forensics approach for a motion compensated video known as the "Game Theoretic Framework". The purpose was to identify the optimal set of actions for forensics investigators and forgers alike. Their simulation results revealed that any false-alarm constraint is less or equal to 10\%. Moreover, forensics investigators have a 50\% chance of detecting a video forgery, and in case the false alarm constraint was higher than 15\%, the detection rate of video forgeries was equal to or higher than 85\%.
  In~\cite{baier2014afauc}, Baier et al. presented an approach called Anti-Forensics of storage devices by alternative Use of Communication channels (AFAUC), which relies on reverse engineering of the firmware commands to access a storage medium through the communication channel. In fact, the approach can be achieved without expensive toolkits, with a lower risk of detection. In~\cite{stamm2012temporal}, Stamm et al. presented an approach based on the use of a game theoretic framework to identify the optimal set of actions for both the forensics investigator and the forger model. This helped them design an anti-forensics technique with the ability to remove any frame fingerprint by deletion or addition. Moreover, the authors showed that their presented anti-forensics technique can fool forensics techniques if applied at full strength.\\
 In~\cite{shirani2002anti}, Shirani et al. aim at hiding the intrusion attempt. In~\cite{peron2005digital}, Peron et al. aimed to limit the collection, identification, and validation of electronic data. Garfinkel~\cite{garfinkel2007anti} and Rogers~\cite{rogers2006anti} aimed to defeat any forensics analysis by limiting the quantity and quality of forensics evidence. In~\cite{rogers2005anti}, Foster and Liu managed to evade and avoid detection by leading anti-forensics attacks. In~\cite{dahbur2013toward}, Dahbur et al. presented the use of scientific methods to confuse the forensics investigation in all its stages. In~\cite{albano2011novel}, Albano et al. presented different methods to thwart a given digital investigation process. In~\cite{sremack2007taxonomy}, Sremack and Antonov presented a taxonomy to thwart a forensics investigation. In~\cite{stamm2012forensics}, Stamm et al. managed to disguise and manipulate, and falsify the devices specific fingerprints once a digital file is formed.

\begin{figure*}[!ht]
  \centering
  \begin{minipage}[b]{1\textwidth}
    \includegraphics[width=\textwidth]{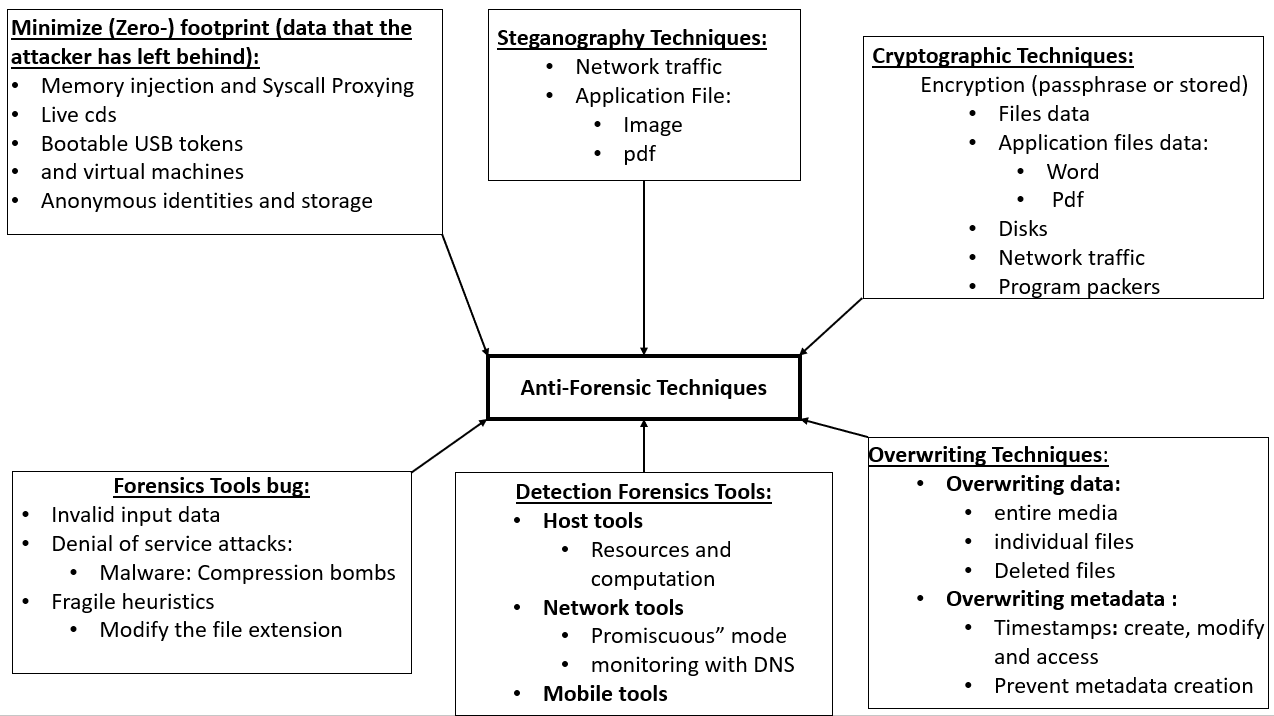}
    \caption{Anti-Forensics Techniques} % HN: I like it
    \label{fig:k}
  \end{minipage}
\end{figure*} 
 
\par
Anti-Digital Forensics (ADF) consists of identifying any activity aiming to hide an attack's trace(s)~\cite{dekker2008taxonomy}. ADF is used by forensics investigators, forensics researchers, and first responders. Their main anti-forensics techniques~\cite{kessler2007anti} are summarized in \figurename~\ref{fig:k}, and are classified as follows:
\begin{itemize}
\item \textbf{Hiding Data:} 
 Encryption and steganography~\cite{channalli2009steganography,kumar2010steganography} are mainly used to hide any evidence and cover criminals tracks to extremely complicate a forensics investigation. This includes encrypting data, encrypting disks, hiding data in the network traffic or even in the memory, etc.
\item \textbf{Encrypting Data:}
Encrypting data is the straightforward way to hide data from being easily disclosed. Encryption also prevents any unauthorised access to the stored data. Cyber-criminals use the encryption to make it harder to investigate and extract data. Therefore, leading to a total waste of time and resources to decrypt long keys that encrypt false data. %In addition, encryption and steganography can be combined together to extremely complicate a forensics investigation.
\item \textbf{Secure-Deletion:}
Secure-Deletion consists of removing the targeted data completely and permanently from the source system by overwriting it with random data. This ensures that the data will not be recoverable anymore. However, most commercial secure-deletion tools do not ensure a full deletion of data, as some parts of it might still be recovered~\cite{srinivasan2007security}.
\item \textbf{Hashing:}
is used by criminals to evade detection by preventing the validation of data's integrity. In this regard, various techniques were used including, fuzzy hashing~\cite{breitinger2012fuzzy}, hash collision, MD5~\cite{rivest1992md5} and SHA-1~\cite{eastlake2001us}.The hash's generated output is unique and can be used as a biometric print for a given input file. Therefore, in case of a minor change in the original file, the hash value is completely different. Resulting into the recovery of the original input file to become almost impossible. %As a result, the recovery of the original input is almost impossible.
\item \textbf{Encrypting Disks:}
 Different tools were developed to encrypt the full hard-drive’s volume. Thus, cyber-criminals employ disk encryption to protect any data that may serve as evidence against them. This can be done by converting it into an unreadable non-comprehensive form, or unsupported format. Making it difficult for digital forensics investigators to decipher it. Moreover, disk encryption relies on encrypting software or/and hardware to encrypt every bit of data that exists on the hard disk~\cite{wasilewski2009retrieval}.
\item \textbf{Encrypting Databases:}
Due to the constant increase use of databases~\cite{davida1981database}, database encryption became another popular form of data hiding. This encryption also targets single users and multi-users' files/ folders. Database encryption~\cite{agrawal2004order} is based on the process of converting data into a meaningless cipher text, including applications, emails, mobile devices, and cloud services.
\item \textbf{Hardware Memory Encryption:} the development towards this type of memory encryption helps criminals avoid access hierarchies of a traditional memory. This will render any known memory acquisition form as infeasible~\cite{latzo2019universal}.
\item \textbf{Steganography:} 
Cyber-criminals use steganography to hide data into digital multimedia elements. These elements include image, video, audio, and text files~\cite{bergmair2004natural,lubacz2010vice}. This also includes system files, as presented by Peron et al. in~\cite{peron2005digital}. Steganography can be overcome by relying on steganalysis methods and attacks \cite{meghanathan2010steganalysis,kaur2014review,ge2011steganography}.
\item \textbf{Data Contraception:} 
This method was introduced by Conlan et al.~\cite{conlan2016anti}, as a new way for hiding data. In fact, it is classified as an anti-forensics activity that leaves either little or no traceable digital evidence(s) to prevent its retrieval. In fact, data contraception can deliberate file-systems and manipulate hard-disks in use by hiding any item on a given system or network.
\item \textbf{Zero-Footprinting:}
 or disk cleaner is a new emerging anti-forensics tool~\cite{sartin2006anti} which is used to clean disk areas or completely destroy the disk’s original content(s). Thus, making the attack completely undetectable. Zero-footprinting shows its usefulness for legitimate or/and illegitimate purposes, due to its ability to un-link files and overwrite them with gibberish data.
\item \textbf{Timestamp Modification:}
or timestamp extraction is a critical task that requires an establishment of a forensics chain-of-events investigation. However, hackers and cyber-criminals managed to modify timestamps of files and logs to mislead investigators. For additional details, different timestamp modification tools are already mentioned in~\cite{wang2005break}.
\item \textbf{File Signature Manipulation:}
A file signature exists at the beginning of each file to identify a given file type. Hackers usually use anti-forensics tools to purposely change and manipulate a file signature to mislead forensics investigators~\cite{dahbur2011anti}.
\item \textbf{Hiding Network:}
Networks were also used by the attackers to hide data. The aim  of hiding data into networks is to ensure that no traces are left behind the attackers. Thus, resulting into crippling the forensics investigation, especially due to the use of VPNs, proxies or TOR.
\item \textbf{Artefact Wiping:}
Artefact Wiping~\cite{jain2014anti} consists of the destruction of useful data that serves as a possible evidence~\cite{harris2006arriving}. Through the analysis of artefact wiping, many software tools exist and can be used to wipe different forms of data and metadata. This includes files, disks, logs, audits and registers. In fact, various types of tools were built by combining different data wiping forms.
\item \textbf{Trail obfuscation:} 
is a deliberate activity to purposely disorient and divert a forensics investigation. It is based on the same principles of steganography, or false data injection~\cite{botas2015counterfeiting}. Trail obfuscation employs the Peer-to-Peer (P2P) protocols to perform cyber-criminal activities. This helps cyber-criminals to mitigate their cyber biometric “fingerprints” to hide the evidences and cover up their tracks. 
\item \textbf{Virtual System Execution:}
The execution of a malicious code or script can be either led from external or even remote disk storage without leaving any trace(s) on the device. Moreover, in~\cite{botas2015counterfeiting}, Botas et al. presented different virtualization mechanisms including USB boot devices and network boot devices. 
\item \textbf{Content Compression:}
 Saturation or content compression is aimed at infecting systems with unstable contents. This leads to added latency and delays, which have a high negative impact on the forensics investigation process. Content compression can be divided into two types: \textbf{Compression Bombs} or zip bombs~\cite{botas2015counterfeiting}, are designed to largely expand once decompressed, and \textbf{Regular Compression} used to exploit regular compression implementations~\cite{chen2016dispersing}.
\item \textbf{Data Pooling:}
 By data pooling, attackers intend to keep their digital media including USB keys, CDs/DVDs, smartphones, laptops, PCs and hard drives active. By doing so, investigators are lured to search all the collected data. As a result, such a search can take from months to years, and may violate the suspect's or victim's privacy, which would result into legal conflicts \cite{aleneziiot}. Therefore, leading to a higher investigative cost, and longer investigation time. 
\item \textbf{Loop References:}
are identified as default file paths lengths, which are restricted to 260 characters due to the Windows Application Programming Interface (API) on New Technology File System (NTFS). However, various ways to initiate longer paths exist. The most popular way is based on the use of Long Path Tool (LPT)~\cite{gul2017survey}. Other ways also exist, including the use of loop references where symbolic links can point to a parent folder. Thus, creating a recursive path, where malicious users can safely store their data in these recursive nested files.
\item \textbf{Dummy Hard Disk:}
Hackers and cyber-criminals use this method by keeping an unusable PC with a hard disk. This allows the PC to be booted from a USB where the OS is stored, without using the hard disk itself. Therefore, data will be stored on cloud services. Hackers might also try and simulate random writes on hard disks, to trick investigators into thinking that a given hard disk has been recently used~\cite{perklin2012anti}. Therefore, this would result into a waste of time and resources.
\item \textbf{Anti-Forensics Malware:} which were also used to perform an anti-forensics activity by wiping out all relevant data that serve as a vital evidence to track down its source, structure and characteristics. Among these malware types we name Stuxnet 1-2, Duqu, Duqu 2.0, Flame, Red October, Shamoon, Gauss, and Mahdi malware types \cite{collins2012stuxnet,bencsath2012duqu,zhioua2013middle,villeneuve2013operation,zaored,dehlawi2013saudi}, used for cyber-warfare \cite{white2018understanding}, cyber-terrorism \cite{lewis2002assessing}, cyber-politics (hacktivism) \cite{jordan2004hacktivism,applegate2011cybermilitias} and cyber-(industrial)-espionage \cite{buttonindustrial} purposes.
\end{itemize}

\section{Anti-Anti-Forensics}
~\label{sec:5}

Categorizing anti-anti-digital forensics includes the classification, identification, characterization , and the differentiation between digital forensics and anti-forensics techniques and tools~\cite{harris2006arriving}. In~\cite{peron2005digital}, the effectiveness level of anti-anti-forensics tools against the use of the old fashion traditional anti-forensics tools was evaluated.
As a result, different anti-anti-forensics approaches were presented.

\subsection{Anti-Forensics Prevention Techniques}
Anti-forensics preventing techniques were presented to counter the anti-forensics activities whilst also preserving the privacy of both individuals and evidences. One key challenge in digital forensics is to protect the privacy of the digital evidences~\cite{benjamin2010yu,dehghantanha2014privacy} during a forensics investigation~\cite{andl2004epithelial,law2011protecting}. Thus, several forensics solutions have been developed to preserve the privacy of the evidences, including digital files, emails or even documents.  In~\cite{goh2003secure}, Goh et al. presented a secure indexing scheme used to search for encrypted data and support advanced query searches~\cite{song2000practical}. This also included the Hash Message Authentication Code (HMAC) and Advanced Encryption Standard (AES) stream cipher operations to ensure a high level of accuracy and efficiency, while guaranteeing the admissibility of electronic evidence and the privacy of each individual. In~\cite{stahlberg2007threats}, P. Stahlberg et al. investigated the privacy threats that may possibly surround the database investigation and proposed a system transparency criteria set. This system set is used to control the results of different queries, except for database searching and retrieval. In~\cite{bottcher2008detecting}, S.Bottcher et al. presented a detective database forensics approach to be capable of detecting any privacy leakage. This was done through the identification of each party accessing the leaked information. In~\cite{reddy2009forensic}, Reddy et al. presented a theoretical forensics readiness framework, which can be used exclusively for enterprises and organisations. This framework suggested a specific organisational structure used to minimize the risk of possibly leaking private information in a given digital investigation case. In~\cite{guo2010research}, Guo et al. defined general policies and procedures for network forensics investigations. In~\cite{pangalos2010importance}, Pangalos et al. provide a description of the forensics readiness role when it comes to optimizing the level of security and privacy of each organisation. In~\cite{croft2010sequenced}, N.J. Croft et al. presented a sequential private data release model which is based on the prior knowledge and proof of a given hypothesis used for forensics investigations. This resulted into placing the less important data in less sensitive layers, allowing sensitive and important data to be made available only in case of the knowledge of lower-level layers. This process was proven and demonstrated by the forensics investigators. In~\cite{law2011protecting}, Law et al. presented several cryptographic models which can be employed into the already existing digital forensics processes to ensure a higher level of data protection. In~\cite{pearson2011privacy}, S. Pearson developed a privacy model and language which can be incorporated within a given company. This helps ensuring auditing and assurance of the employed mechanisms. In~\cite{pooe2012conceptual}, Pooe et al. studied a forensics policy specification to ensure a higher forensics readiness. In~\cite{hou2011privacy}, S.Hou et al. investigated the legal and practical privacy issues in a given forensics investigation and successfully presented a practical solution based on using homomorphic and communicative encryption techniques to limit the disclosure of data during a given forensics investigation. However, their solution lacked the ability to identify malicious data from non-malicious data~\cite{lin2010efficient}. In~\cite{gupta2013privacy}, Gupta presented a a framework called “Privacy Preserving Efficient Digital Forensics” (PPEDF) to ensure an automated investigation through the reduction of the amount of data being analyzed. In fact, PPEDF is compatible with the Encase Version 7.0, with a 100\% accuracy when extracting evidence files. In~\cite{hou2013privacy}, Hou et al. presented another solution based on the use of the (t,n) sharing scheme as a data encryption method, to ensure the integrity and authenticity through the appliance of homomorphic property of the (t,n) sharing scheme. In~\cite{armknecht2015privacy}, Arrnknecht et al. presented another privacy preserving mechanism for email data. This method is based on the combination of secret sharing and encryption algorithms. In fact, it is based on two main schemes including protection and extraction. This ensures that the encrypted data will only be decrypted upon its need. In~\cite{afifah2016development}, Afifah et al. revealed and alternative implementation of data protection presented by Armknecht and Dewald. It was focused on preserving the privacy of disk image instead of email data. In~\cite{nieto2018iot}, Nieto et al. presented a solution named \textbf{Digital Witness}, which is a personal device that identifies, collects, safeguards and communicates digital evidences~\cite{nieto2016digital} as a member of Digital Chains of Custody in Internet of Things (DCoC-IoT)~\cite{prayudi2015digital}. This was meant to support eleven privacy principles included in several PRoFIT (Model The Privacy-aware IoT-Forensics) presented in~\cite{nieto2017methodology}. Thus, ensuring a better cooperation between citizens and digital forensics investigations. 

\subsection{Anti-Forensics Detection Techniques}
Anti-Anti-Forensics is a newly evolving technology that protects forensics against any anti-forensics attempt(s). Hence, it is essential to maintain the right anti-forensics countermeasures to ensure a high detection rate of any anti-forensics activity or attack. 

The file installation for cryptographic software indicated that the data could be possibly encrypted on a system which could lead to the occurrence of a possible anti-digital forensics activity. Therefore in~\cite{conlan2016anti}, Conlan et al. compared a hash data set against NIST hashes, where unmatched hashes were possibly a sign of the existence of anti-forensics files or/and tools. This indicated the possibility of the employment of anti-digital forensics tools to erase any evidence beyond recovery to cover all tracks. Disk-avoiding using anti-forensics tools was addressed by the Garfinkel in~\cite{garfinkel2007anti}. The presented solution is built on existing anti-forensics detection methods. In~\cite{blunden2009anti}, Blunden examined the already existing approaches that might be used by a forensics investigator against the malicious yet persistent use of rootkits, whilst identifying the anti-forensics possibilities that a rootkit might use or even employ. 
\\
To mitigate the use of anti-forensics activities, an enhanced protected forensics version is needed. Hence, the shifting is heading towards an enhanced Anti-Anti-Forensics version~\cite{perklin2012anti}.
In~\cite{conlan2016anti}, a theoretical approach was presented by Conlan et al. to detect the use of anti-digital forensics tools, and reporting them to digital investigators. This enhanced and improved  the digital forensics investigation to overcome an anti-forensics attack~\cite{rekhis2012system}. 
In~\cite{geiger2005evaluating,geiger2005counter}, Geiger presented an approach that consists of the analysis of five  anti-forensics tools including: “Secure-Clean”, “Evidence Eliminator”, “Window Washer”, “Cyber-Scrub Professional”, and “Acronis Privacy Expert”. This was done by using a Forensics Tool Kit (FTK). The approach revealed that an incomplete wiping of unallocated space allowed the recovery of the right data containing the necessary evidences. In~\cite{fairbanks2007timekeeper}, Fairbanks et al. introduced a forensics tool, that can extract and analyse the forensics data to allow the detection of anti-forensics attempts, whilst also capturing unavailable forensics information. This facilitated a system’s recovery, and enhanced the digital forensics investigation.
\\
Despite the challenges and limitations that forensics domains suffer from, machine learning came as a new early smart detection method to sort the limitations of previous forensics and counter anti-forensics methods. As a result, a new machine-learning counter anti-forensics-based branch was presented in~\cite{rughani2017machine,hoelz2009artificial,mitchell2010use,allen2017artificial} to detect any anti-forensics activity. In~\cite{conti2018cyber} Conti et al. revealed the importance of implementing and applying Artificial Intelligence-Machine Learning (AI-ML) techniques in the cyber-security domain. In \cite{mukkamala2003identifying}. Mukammala et al. presented a study based on the use of artificial intelligent techniques (Artificial Neural Networks (ANNs) and Support Vector Machines (SVMs)) for offline intrusion analysis to maintain the integrity and confidentiality of the information infrastructure. Based on their study, SVM outperformed ANN in terms of scalability and prediction accuracy, while both methods produce largely consistent results. In \cite{yeow2014application}, Yeow et al.  designed and developed an Intelligent Forensic Autopsy (of war victims) Report System (I-AuReSys)  based on the Case-Based Reasoning (CBR) method, which is used to analyse forensic evidence. I-AuReSys is used to extract features by using an information extraction (IE) technique from the already existing autopsy reports, before analysing any case similarities by coupling CBR technique with a Naïve Bayes learner for feature-weights learning. Experimental results class as a practical viable alternative forensics method. In \cite{weber2019anti}, Weber et al. presented graph convolutional networks for financial forensics as prototype and experimented them to overcome  Anti-Money Laundering (AML) in bitcoin cryptocurrency. The authors contributed to the Elliptic Data Set using series graph of Bitcoin transactions (nodes), directed payment flows (edges), and node features, including ones based on non-public data. Results revealed the superiority of Random Forest (RF) algorithm. In \cite{wang2019tkrd}, Wang et al. presented a novel TKRD method named Trusted Kernel Rootkit Detection for cyber-security of Virtual Machines (VM). TKRD is based on machine learning and memory forensic analysis and is used to detect kernel rootkits in VMs from private cloud. Experimental results revealed that the RF classifier has the best unknown kernel rootkits detection performance. In \cite{axenopoulos2019framework}, Axenopoulos et al. presented a new framework which is implemented in the context of the European Union-funded project LASIE. This framework is applied for large-scale exploitation of forensic data acquired from different sources and in multiple formats, whilst several video analytics tools that performed automated object (human, face, vehicle, logo) detection and tracking, video event detection and summarization. Detection and tracking events are also robust in low-resolution, low color quality, motion blur, and lighting variations. An evidence search engine was also presented to offer various ways of retrieving relevant evidence. This framework was tested using real content (CCTV footage) provided by the London Metropolitan Police (MET), and have shown promising results. In \cite{sun2018novel}, Sun et al. presented a novel Convolutional Neural Network-based (CNN-based) Contrast Enhancement (CE) forensics method, using the the Gray-Level Co-occurrence Matrix (GLCM) which contains traceable CE forensics features. Experimental results revealed that this method outperforms conventional forensics methods in terms of forgery-detection accuracy, robustness and performance, especially when dealing with counter-forensic attacks. In \cite{cao2018robust}, Yang et al. presented two effectively robust CE forensics algorithms based on deep learning. Their method achieves a better end-to-end classification based on pixel and histogram domain. Experimental results revealed that this method achieves a better detection performance than the other state-of-the-art algorithms, as well as it being robust against Pre-JPEG compression and anti-forensics attacks. In \cite{shan2019robust}, Shan et al. presented a JPEG-robust CE forensic method based on a modified CNN, adding a GLCM layer and cropping layer ahead of a tailor-made CNN. Extensive experimental results revealed that this method achieves significant improvements in terms of both global and local CE detection. In \cite{yu2016multi}, Yu et al. presented a multi-purpose CNN-based method to detect various anti-forensics activities, including the automatic features extraction and identification of the forged types. This model can effectively detect various image anti-forensics in binary and multi-class decision. Experimental results revealed that their methods achieves a better performance than other counter-anti-forensics methods in anti-forensics detection. In \cite{chen2018densely}, Chen et al. presented a new CNN approach for multi-purpose image detection and manipulations under anti-forensics activities, and using a dense connectivity pattern for a better parameter efficiency. Experimental results revealed a better performance in terms of accuracy detection of anti-forensics attacks, as well as enhanced robustness against JPEG compression. In \cite{li2018learning}, Li et al. presented a 3D Convolutional Neural Network architecture tailored for the spatial-temporal input to tackle the face spoofing detection problem. Experimental results revealed that this method can learn discriminative and generalized information compared to with other deep learning based biometric spoofing detection methods.

\par
In the next section, the main recommendations will be presented to ensure a much more suitable forensics solution to adhere and overcome various forensics challenges and issues especially in terms of security and privacy-wise.

\section{Suggestions \& Recommendations}
~\label{sec:6}

Due to the increase number of data volume, size, structure and velocity, this paper suggests and recommends the following solutions:
\begin{itemize}
\item \textbf{Counter Anti-Forensics:} further work needs to be done in terms of ensuring a higher accuracy and detection rates based on employing further machine-learning-based approaches, whilst also employing different privacy preserving solutions to prevent any evidence alteration caused by anti-forensics activities.

\item \textbf{Enhancing the investigators skills:} investigators must be legally certified by undergoing further training to specialise in the cyber-security and digital forensics fields. As well as be more familiar with how to use forensics/counter anti-forensics techniques and tools, to enhance their investigation and investigative skills. This will make it less complex and less time consuming~\cite{laykin2013investigative}. 

\item \textbf{Forensics Training/Testing Ground:} is required and more funding is needed especially for the newly emerging forensics tools. This would help ensure their accuracy, advantages, limitations and issues through forensics testing. Moreover, forensics and digital forensics examination and educational grounds (low level/high level courses) need to be reconsidered and re-evaluated to adhere to the modern constant growth in this domain.

\item \textbf{Raising Forensics Awareness:} can be enhanced through constant workshops, and forensics-based events, as well as weekly, monthly or yearly meetings and international conferences.

\item \textbf{Constant Alertness \& Awareness:} is required in order to monitor and shadow the newly or/and constantly emerging topics, where forensics can play a key role to locate, identify, retrieve and protect evidences. This can be done by expanding the range of forensics fields to cover every digital, real-life and IoT aspects.

\par

\end{itemize}

\section{Conclusion}
~\label{sec:7}

The integration of forensics into the digital field and the IoT world led to its global spread and worldwide use to sort digital, cyber and IoT-related crimes. However, in recent years, there was a very remarkable rise in the number of anti-forensics activities to hide evidences and alter/delete them beyond recovery. In this paper, a new modern forensics analytical view is presented. An initial forensics background was presented to include the forensics investigation process, chain-of-custody and the structure of cyber crimes, while also classifying digital data and digital investigators types. Then, digital forensics sub-domains were discussed along their different investigative forensics tools, techniques, and approaches. Cyber-forensics challenges were also mentioned and detailed. Anti-forensics aspects and techniques were also highlighted, whilst counter anti-forensics detection and prevention techniques were discussed using machine learning techniques to enhance the detection rate and accuracy, whilst using privacy preserving techniques to prevent any evidence alteration, deletion or/and modification.
\\
As part of future work, further research will be performed on the newly introduced counter anti-forensics or anti-anti-forensics topic especially in terms of enhancement in both detection, prevention, and privacy preserving aspects.

\section*{Acknowledgement}
 This paper was partially supported by funds from the Maroun Semaan Faculty of Engineering and Architecture at the American University of Beirut.

\bibliographystyle{unsrt}

\end{document}